\documentclass[a4paper,11pt]{article}
\pdfoutput=1

\usepackage{jcappub}

\usepackage{bm}
\usepackage{enumerate}
\def\dd{\mathrm{d}}

\def\equationautorefname~#1\null{%
  eq.~(#1)\null
}
\def\figureautorefname~#1\null{%
  figure~#1\null
}
\def\tableautorefname~#1\null{%
  table~#1\null
}

\title{\boldmath
  Isocurvature bounds on axion-like particle dark matter
  in the post-inflationary scenario
}

\author[a,b,c]{M. Feix,}
\author[d]{J. Frank,}
\author[c,d]{A. Pargner,}
\author[e]{R. Reischke,}
\author[b,c]{B.M. Sch\"afer,}
\author[c,d]{T. Schwetz}

\affiliation[a]{Institut f{\"u}r Theoretische Astrophysik, Zentrum f{\"u}r Astronomie der Universit{\"a}t Heidelberg,\\ Philosophenweg 12, 69120 Heidelberg, Germany}
\affiliation[b]{Astronomisches Rechen-Institut, Zentrum f{\"u}r Astronomie der Universit{\"a}t Heidelberg,\\ Philosophenweg 12, 69120 Heidelberg, Germany}
\affiliation[c]{HEiKA -- Heidelberg Karlsruhe Research Partnership, Heidelberg University,\\ Karlsruhe Institute of Technology (KIT), Germany}
\affiliation[d]{Institut f\"ur Kernphysik, Karlsruher Institut f\"ur Technologie (KIT),\\ Hermann-von-Helmholtz-Platz 1, 76344 Eggenstein-Leopoldshafen, Germany}
\affiliation[e]{Department of Physics, Israel Institute of Technology -- Technion,\\ 3200003 Haifa, Israel}

\emailAdd{feix@uni-heidelberg.de}
\emailAdd{r.reischke@campus.technion.ac.il}
\emailAdd{andreas.pargner@kit.edu}

\abstract{
We assume that dark matter is comprised of axion-like particles (ALPs) generated by
the realignment mechanism in the post-inflationary scenario. This leads to isocurvature
fluctuations with an amplitude of order one for scales comparable to the horizon at the
time when the ALP field starts oscillating. The power spectrum of these fluctuations
is flat for small wave numbers, extending to scales relevant for cosmological observables.
Denoting the relative isocurvature amplitude at $k_* = 0.05$~Mpc$^{-1}$ by $f_{\rm iso}$,
Planck observations of the cosmic microwave background (CMB) yield $f_{\rm iso} < 0.31$
at the $2\sigma$-level. This excludes the hypothesis of post-inflationary ALP dark matter
with masses $m_a < 10^{-20}$--$10^{-16}$~eV, where the range is due to details of the ALP
mass-temperature dependence. Future CMB stage IV and 21-cm intensity mapping experiments
may improve these limits by 1--2 orders of magnitude in $m_{a}$.
}

\keywords{axions, cosmological parameters from LSS, dark matter theory}

\begin{document}
\maketitle
\flushbottom

\section{Introduction}
\label{sec:intro}
Over the last decades, the increasing wealth of astronomical data has firmly established the standard cosmological paradigm where the observed cosmic structure
has formed out of small perturbations in the primordial energy-matter density field and their subsequent growth due to gravitational interaction. Interpreted
within general relativity (GR), the gathered evidence for a mass deficit of visible matter on astrophysical and cosmological scales \citep{Bertone2005, Bertone2018}
as well as strong indications for an accelerated expansion of the Universe \citep{riess_observational_1998, perlmutter_measurements_1999, planck_collaboration_planck_2016}
have led to the now widely accepted cold dark matter (DM) model with a cosmological constant, usually denoted as the $\Lambda$CDM model \citep[e.g.,][]{bartelmann_dark_2010}.
Originally invoked to explain the observed dynamics of galaxies within the Coma cluster \citep{Zwicky1933, Smith1936}, DM remains merely a postulate of the cosmological
model and its true nature is unarguably one of the largest puzzles in modern physics.

A well-motivated particle physics candidate for DM is the axion \citep{Weinberg1978, Wilczek1978, Sikivie2008, Marsh2016} which appears as a pseudo-Nambu-Goldstone
boson (PNGB) in the Peccei-Quinn (PQ) solution to the strong $\mathrm{CP}$ problem \citep{Peccei1977}. Their solution introduces a new global chiral $U(1)$ symmetry
that gets spontaneously broken at some energy scale $f_{a}$ by the vacuum expectation value of a complex scalar field. The axion then emerges as the phase degree of
freedom of this complex scalar. To satisfy current experimental bounds, $f_{a}$ can be assumed very large. This gives the axion very weak interactions and an extremely
small mass \citep{Dine1981, Zhitnitsky1980, Kim1979, Shifman1980}. There exist many other high-energy extensions of the standard model that also contain PNGBs sharing
properties similar to those of the axion. These are commonly referred to as axion-like particles (ALPs) \citep{Arvanitaki2010, Arias2012, Ringwald2012}.

Despite their low masses, axions and ALPs can effectively mimic the properties of a (cold) DM component, thanks to the possibility
of non-thermal production via the vacuum realignment mechanism \citep{Dine1983, Preskill1983, Turner1983, Abbott1983}. At early times
when the temperature $T\sim f_{a}$, the axion (or ALP) field is essentially massless. Since no specific field value is energetically
favoured in this case, it simply takes a random one. At lower temperatures around the QCD phase transition, however, its potential becomes
important, causing the axion field to roll down to the $\mathrm{CP}$ conserving minimum and, therefore, realign with the vacuum. The
energy stored in the coherent oscillations around this minimum then largely behaves like collisionless DM on cosmologically relevant
scales. Considering the QCD axion, a typical mass suitable to explain the observed DM density is $m_{a}\sim 10^{-5}$ eV
\citep{Visinelli2009, Kawasaki:2014sqa, Klaer2017}. For ALPs, there is substantially more freedom in the allowed mass range which may
extend to much smaller values \citep[e.g.][]{Arias2012}.

There exist two fundamentally different scenarios depending on how the vacuum realignment mechanism is realized in the early Universe.
If the PQ symmetry is broken before (or during) an inflationary epoch, the axion (or ALP) field takes a single value in the whole observable
Universe that sets its energy density in an evenly distributed manner. However, if PQ symmetry breaking happens after the end of inflation,
the situation becomes more involved. In this case, the axion field generally takes different values in causally disconnected regions, giving
rise to large isocurvature fluctuations in the axion energy density field. This has interesting phenomenological consequences for axion DM,
in particular the formation of gravitationally bound objects known as miniclusters \citep{Hogan1988, Kolb1993, Kolb1994b, Kolb1996, Zurek2007,
Hardy2016, Enander2017, Vaquero2018}.

In contrast to the QCD axion, the presence of ultra-light ALPs (ULAs) can have a sizable and potentially measurable effect on cosmological
large-scale observables. For instance, observations of the cosmic microwave background (CMB) anisotropy spectrum and large galaxy surveys
have already been used to constrain ULAs, essentially excluding ULA masses within the range $10^{-32}$ eV $< m_{a}< 10^{-25}$ eV \citep{Amendola2006,
Marsh2010, Marsh2012, Hlozek2015, Hlozek2018}.\footnote{Note that ULAs with masses $m_{a}\lesssim 10^{-27}$ eV cannot account for all of the
observed DM in the Universe since field oscillations would commence only after matter-radiation equality, and axions would mainly manifest
as a dark energy component.}
Considering smaller (nonlinear) scales, comparisons of the observed Lyman-$\alpha$ flux power spectrum with predictions based on hydrodynamical
simulations suggest an increased lower bound of $m_{a}\gtrsim 10^{-21}$ eV \citep{Haehnelt2017, Kobayashi2017}. Even tighter constraints may
be obtained by introducing additional assumptions, for instance, from the formation of solitonic cores in DM halos \citep{Marsh2018} or from
the spin-down of super-massive black holes via superradiant instability \citep[e.g.,][]{Arvanitaki2015, Stott2018}.

While it is well-known that observational bounds on isocurvature fluctuations in the axion field generated during inflation \citep{Akrami2018a}
constrain the \emph{pre-inflationary PQ breaking scenario} \citep[e.g.,][]{Turner:1990uz, Lyth:1991ub, Beltran2007,Hertzberg2008, Hamann:2009yf, Visinelli2017, Schmitz2018},
we will show in the present work that similar isocurvature constraints also apply in the \emph{post-inflationary PQ breaking scenario}. To this
end, we will consider the more general context of ALPs and assume that the vacuum realignment mechanism is the most relevant production mechanism
of such particles in the early Universe. In addition to adiabatic modes, this scenario generically predicts cosmological isocurvature fluctuations
that are characterized by a much steeper power spectrum than the usually assumed (nearly) scale-invariant spectrum produced during inflation. The
origin of these isocurvature fluctuations is the large inhomogeneity of the ALP field between causally disconnected regions at the time when the
field starts to oscillate and behaves as DM. By looking at their corresponding imprints on CMB temperature anisotropies and the matter power
spectrum, we will be able to place constraints on large parts of the parameter space where such ALPs constitute the cosmic DM, thereby complementing
existing and predicted bounds on ULA masses from current and future large-scale experiments, respectively. In particular, we will focus on
constraints based on Planck data \citep{planck_collaboration_planck_2016, Planck2016_likelihoods}, and provide forecasts for next-generation
CMB \citep[e.g.,][]{PRISM2014, abazajian_cmb-s4_2016} and HI intensity mapping experiments such as the Square Kilometre Array (SKA)
\citep{SKA2015, SKA2018a, SKA2018b}.

This work is structured as follows: in \autoref{sec:axions}, we introduce our model for ALP DM in the context of post-inflationary PQ
breaking. Assuming that ALPs are mainly produced through the vacuum realignment mechanism, we present estimates of the cosmological relic
abundance as well as the initial power spectrum of isocurvature fluctuations. The latter's effect on CMB anisotropies and the matter
power spectrum is discussed in \autoref{sec:cosmology}, and then used to derive strong limits on ULA DM based on current and future
large-scale datasets in \autoref{sec:results}. We conclude and summarize our findings in \autoref{sec:conclusions}.

Throughout, we will assume a spatially flat reference cosmology based on \citep{collaboration_planck_2016}, adopting the total matter density
parameter $\Omega_{\rm m} = 0.315$, the baryon density parameter $\Omega_{\rm b} = 0.049$, the amplitude $A_{\rm s} = 2.215\times 10^{-9}$
of the primordial adiabatic spectrum, its spectral index $n_{\rm s} = 0.9603$ (without running, i.e. $\alpha_{\rm s} = 0$), the optical depth
$\tau = 0.089$, the dimensionless Hubble parameter $h = 0.673$, and the sum of neutrino masses $\sum m_{\nu} = 0.05$ eV. Additional parameters
relevant to our analysis will be introduced and specified below.

\section{ALPs in the post-inflationary symmetry breaking scenario}
\label{sec:axions}

In what follows, we consider the cosmological evolution of ALP fields in the post-inflationary PQ breaking scenario.
Building on semi-analytical results obtained for the QCD axion in \cite{Enander2017}, we adopt the harmonic
approximation for the potential and focus on the vacuum realignment mechanism which constitutes a largely
model-independent way of producing relic axions in the early Universe \citep{Dine1983, Preskill1983, Turner1983,
Abbott1983, Kolb1990}.

A key difference between QCD axions and ALPs is that the latter do not necessarily exhibit a specific relation
between mass, $m_{a}$, and breaking scale, $f_{a}$. Assuming that a potential for the ALP field $a$ is generated
by some exotic strongly interacting sector, we may write
\begin{equation}
V(a) \approx \Lambda^{4}\left\lbrack 1-\cos\left (\frac{a}{f_{a}}\right )\right\rbrack,\qquad
m^{2}_{a} = \left.\frac{\partial^{2}V}{\partial a^{2}}\right\vert_\mathrm{min} =
\frac{\Lambda^{4}}{f^{2}_{a}}\;,
\label{eq:potential}
\end{equation}
where, in analogy to the instanton potential of QCD axions, $\Lambda^{4}$ takes the role of a topological
susceptibility $\chi$ that is generally model-dependent and will be parametrized below. Hence, the ALP is
characterized by two out of the three parameters $f_{a}$, $m_a$, and $\Lambda$.

A crucial ingredient for the cosmological evolution of ALPs is the temperature dependence of its mass. For
the QCD axion, this is fully determined by non-perturbative QCD effects \citep[see, e.g.,][]{Borsanyi2016a,
GrillidiCortona2016}. For general ALPs, however, it depends on the specifics of the mechanism generating
their mass. In the following, we will assume a power law that turns into the constant zero-temperature mass
$m_{a}=\Lambda^{2}/f_{a}$ for low temperatures,
\begin{equation}
m_{a}(T) = \mathrm{min}\left\lbrack\frac{\Lambda^{2}}{f_{a}},\; b
\frac{\Lambda^{2}}{f_{a}}\left (\frac{\Lambda}{T}\right )^{n}\right\rbrack,
\label{Eq:MassParameterization}
\end{equation}
where the parameter $b$ accounts for the possibility that the zero-temperature mass might not exactly be
reached at $T=\Lambda$, but at
\begin{equation}\label{eq:T0}
T_{0} = b^{1/n}\Lambda\;.
\end{equation}
We will consider values in the range $b\simeq 0.1$--10. The parameter $n\in\mathbb{R}_{+}$ controls how
quickly the mass emerges. Choosing $\Lambda = 75.5$~MeV, $b = 10$, and $n = 4$, \autoref{Eq:MassParameterization}
reproduces the $m_a(T)$-behaviour for the QCD axion as obtained in \cite{Borsanyi2016a} to good accuracy.

\subsection{Cosmic evolution}
\label{sec:axions_cosmo}

The evolution of the (real) ALP scalar field in an expanding universe, expressed in
terms of the angular field $\theta(\mathbf{x}) = a(\mathbf{x})/f_{a}$,
is governed by the action
\begin{equation}
S_{\theta}=\int\dd^4x\sqrt{-g}f^{2}_{a}\left\lbrack -\frac{1}{2}\left (\nabla_{\mu}
\theta\right )\left( \nabla^{\mu}\theta\right )-V(\theta)\right\rbrack,
\label{Eq:ThetaAction}
\end{equation}
where $g$ is the determinant of the well-known Friedmann-Lema\^{i}tre-Robertson-Walker spacetime metric $g_{\mu\nu}$
(with signature $+2$), and
\begin{equation}
V(\theta, T) = m^{2}_{a}(T)\left (1 - \cos\theta\right )
\label{Eq:ThetaPotential}
\end{equation}
is the temperature-dependent ALP potential. Variation of \autoref{Eq:ThetaAction} with respect to $\theta(\mathbf{x})$
yields the equation of motion,
\begin{equation}
\ddot{\theta} + 3H\dot{\theta} - \frac{1}{R^{2}}\bm{\nabla}^{2}\theta + \frac{\dd V}{\dd\theta} = 0\;,
\label{Eq:ThetaEOM_NL}
\end{equation}
where $R$ is the cosmic scale factor, $H = \dot{R}/R$ denotes the Hubble parameter, dots correspond to derivatives with
respect to time, and $\bm{\nabla}$ is defined with respect to comoving coordinates. At early times, well before matter-radiation
equality, the contribution of ALPs to the total energy density budget of the Universe is minuscule, and
we will assume that its impact on the cosmic expansion can be neglected.
As usual, the field's energy density, $\rho_{a}$ (and similarly, its pressure) can be computed from the stress-energy
tensor and is given by
\begin{equation}
\rho_{a} = f^{2}_{a}\left\lbrack\frac{1}{2}\dot{\theta}^{2} + \frac{1}{2R^{2}}\left (\bm{\nabla}\theta\right )^{2} + V(\theta )\right\rbrack.
\label{Eq:EnDensGen}
\end{equation}
The periodic form of the potential renders the cosmic evolution nonlinear and severely complicates its analysis. To make
analytic progress, however, we assume that the potential may be approximated as $V(\theta, T)\simeq m^{2}_{a}(T)\theta^{2}/2$.
Although this ignores potentially important effects such as the formation of strings and domain walls \citep{Kibble1976}
or the emergence of very dense objects in the context of miniclusters \citep{Kolb1993, Kolb1994b}, we will obtain useful
estimates for ALP DM on the large scales relevant to cosmological probes. We further discuss this issue in \autoref{sec:axions_topo}.

Within the harmonic approximation, \autoref{Eq:ThetaEOM_NL} expressed in Fourier space reduces to
\begin{equation}
\ddot{\theta}_{\mathbf{k}} + 3H(T)\dot{\theta}_{\mathbf{k}} + \omega^{2}_{\mathbf{k}}\theta_{\mathbf{k}} = 0\;,
\qquad \omega^{2}_{\mathbf{k}} = \frac{k^2}{R^{2}} + m^{2}_{a}(T)\;,
\label{Eq:ThetaEOM_LIN}
\end{equation}
which is reminiscent of the damped harmonic oscillator. At high temperatures, the ALP field is essentially massless. Ignoring
decaying solutions, we see from \autoref{Eq:ThetaEOM_LIN} that $\theta_{\mathbf{k}} = {\rm const}$ for superhorizon modes
with $\omega_{\mathbf{k}} \ll 3H$. Upon entering the horizon, these modes will start oscillating. To characterize this
transition, we introduce $T_{\rm  osc}$ as the temperature where the zero-mode starts to oscillate,
\begin{equation}
3H(T_{\rm osc}) = m_{a}(T_{\rm osc})\;.
\label{eq:T_osc}
\end{equation}
For all modes, the mass term will eventually dominate at late times,
and the ALP energy density given by \autoref{Eq:EnDensGen} will behave
like cold DM. The Hubble rate during radiation domination is given by
\begin{equation}
  H(T) \approx 1.66 \sqrt{g_*(T)} \frac{T^2}{M_P}\;,
\end{equation}
where $g_*$ denotes the effective number of relativistic degrees of freedom and $M_P$ the Planck mass.
With the parametrization from \autoref{Eq:MassParameterization} for $m_a(T)$, we can estimate
\begin{equation}
  T_{\rm osc} \sim  \Lambda \left(\frac{b}{\sqrt{g_*(T_{\rm osc})}}\frac{M_P}{f_a}\right)^{1/(2+n)} \;.
\end{equation}
Using that $\sqrt{g_*(T_{\rm osc})} \simeq$ few and comparing with
\autoref{eq:T0}, we see that for reasonable values of $b$ and $n$, we
have $T_{\rm osc}/T_0 \sim (M_p/f_a)^{1/(2+n)}$. Hence, the condition
$f_a \ll M_P$ implies that the ALP field starts oscillating before
reaching its zero-temperature mass. Therefore, the details of the
temperature dependence are important and our results will depend to
some extent on the parameters $b$ and $n$. For our calculations, we
solve \autoref{eq:T_osc} numerically by interpolating the effective
degrees of freedom tabulated as a function of temperature from
\cite{Borsanyi2016a}, extended to temperatures below
MeV by using the results of \cite{Husdal2016}.

\subsection{Relic density}
\label{sec:axions_relic}

To estimate the cosmological mean density, we note that contributions from modes with $\lvert\mathbf{k}\rvert\neq 0$
will be quickly suppressed relative to the zero mode due to the factor $R^{-2}$ appearing in $\omega^{2}_{\mathbf{k}}$.
Considering only the zero mode (which is equivalent to dropping the gradient terms in the evolution equation), using
a WKB ansatz allows one to obtain an approximate expression for the cosmic mean density valid for temperatures
$T<T_{\rm osc}$ \citep{Dine1983, Preskill1983, Turner1983, Abbott1983, Visinelli2009},
\begin{equation}
\overline{\rho}_{a}(T) \simeq \frac{1}{2}f^{2}_{a}m_{a}(T_{\rm osc})m_{a}(T)
\left\lbrack\frac{R(T_{\rm osc})}{R(T)}\right\rbrack^{3}\langle\theta^{2}_{\rm ini}\rangle\;,
\label{Eq:EnDenRes}
\end{equation}
where $\langle\theta^{2}_{\rm ini}\rangle = \pi^{2}/3$ is the mean
value of $\theta^{2}_{\rm ini}$ averaged over many different Hubble
patches around $T=T_{\rm osc}$. A more rigorous approach is to
directly solve the system given by \autoref{Eq:ThetaEOM_LIN} for all
relevant modes. Rewriting $\theta_{\mathbf{k}}(R) =
\theta^{i}_{\mathbf{k}}f_{\mathbf{k}}(R)$, where
$\theta^{i}_{\mathbf{k}}$ is the initial value at time $t_{i}$ and
$f_{\mathbf{k}}(R)$ captures its evolution with
$f_{\mathbf{k}}(R_{i}) = 1$, the mean energy density can be formally
related to the field's initial power spectrum $P_\theta(k)$,
\begin{equation}
\overline{\rho}_{a} = \frac{f_{a}^{2}}{2}\int_{0}^{\infty}\frac{\dd^{3}k}{(2\pi)^{3}}
\, P_{\theta}(k)F(\mathbf{k},\mathbf{k})\;,
\label{eq:mean_proper}
\end{equation}
where $\theta(\mathbf{x})$ is assumed as statistically homogeneous and isotropic, $P_\theta(k)$
is defined through $\langle\theta_{\mathbf{k}}\theta^{\ast}_{\mathbf{k}^{\prime}}\rangle =
(2\pi )^{3}\delta_{D}(\mathbf{k}-\mathbf{k}^{\prime})P_\theta(k)$, and
\begin{equation}
F(\mathbf{k},\mathbf{k}^{\prime}) = \dot{f}_{\mathbf{k}}\dot{f}_{\mathbf{k}^{\prime}} +
\left\lbrack\frac{\mathbf{k}\cdot\mathbf{k}^{\prime}}{R^{2}} + m_{a}(T)\right\rbrack f_{\mathbf{k}}f_{\mathbf{k}^{\prime}}\;.
\label{eq:kernel}
\end{equation}
The expression in \autoref{eq:mean_proper} is of the same parametric form as \autoref{Eq:EnDenRes},
but replaces $\langle\theta^{2}_{\rm ini}\rangle = \pi^{2}/3$ with a proper weighted contribution
of non-zero modes specified by $P_{\theta}(k)$ \citep{Enander2017}.

In the post-inflationary scenario, the ALP field $\theta$ will assume
uncorrelated values in causally disconnected regions whereas the
gradient terms in the evolution equation tend to equalize the value of
$\theta$ inside the horizon \citep{Kibble1976} (see, e.g.,
\cite{Gorghetto2018} for recent simulations). This can be described
by an initial power spectrum $P_{\theta}(k)$ corresponding to a
constant (i.e. white noise) for $k < R H$, whereas power for $k
> R H$ is suppressed. Here ``initial'' means shortly before the
ALP field starts to oscillate. Following \cite{Enander2017}, we choose
an exponential shape for the power spectrum, $P_\theta(k) \propto
\exp(-k/Q)$ with $Q = R_i H(T = T_i)$, and set the initial
time to $T_i = 3 T_{\rm osc}$. The amplitude of the power spectrum for
$k \ll Q$ is fixed by requiring that $\langle\theta^2\rangle =
\pi^2/3$.

To determine $f_\mathbf{k}(t)$, we approach the equation of motion
numerically. We use a realistic temperature dependence of $g_{\ast}$ from
\cite{Borsanyi2016a, Husdal2016}, and solve \autoref{eq:T_osc} for
temperatures from $T_{i}$ down to a few times less than $T_{\rm osc}$.
This solution is then matched onto a WKB approximation for lower temperatures
to factor out the fast oscillations of the axion field (see \cite{Enander2017}
for details). For a given temperature dependence of the ALP mass, we can
calculate the energy density using \autoref{eq:mean_proper}.

\begin{center}
\begin{figure}
\includegraphics[width=0.32\textwidth]{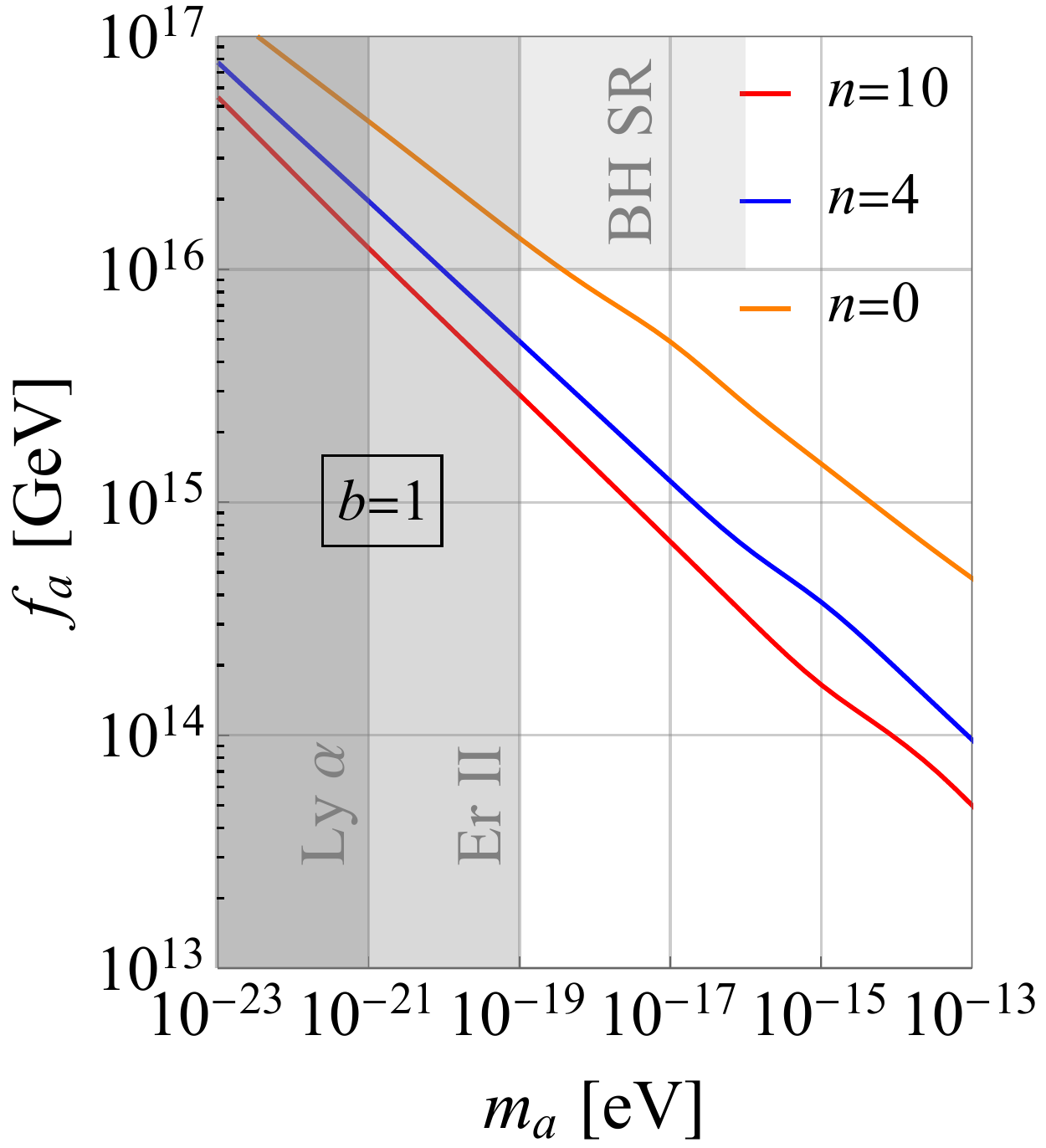}
\includegraphics[width=0.322\textwidth]{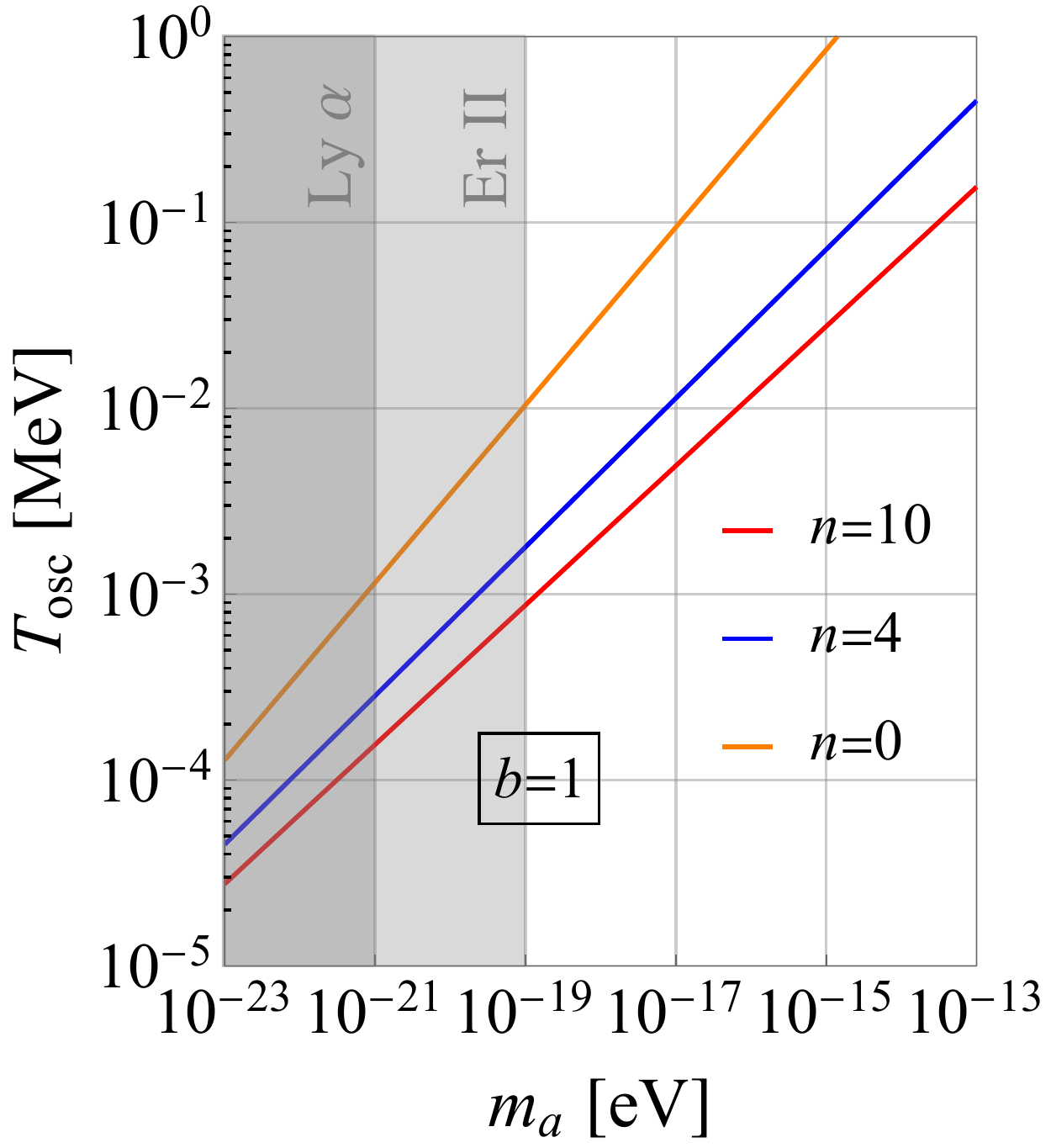}
\includegraphics[width=0.312\textwidth]{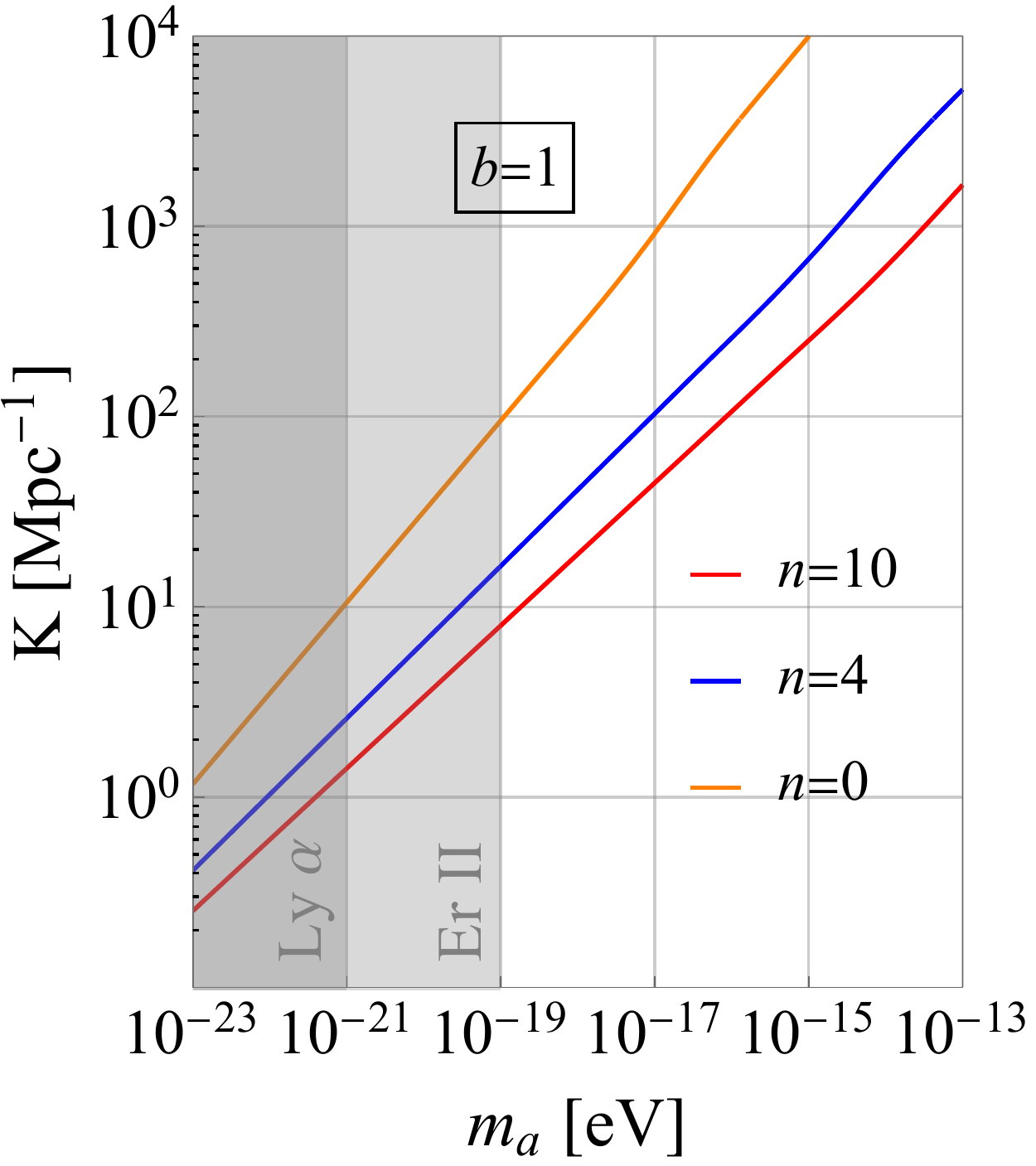}
\caption{The left panel shows $f_{a}$-values that yield the correct DM abundance according to
  \autoref{eq:mean_proper} as a function of $m_{a}$. These $f_{a}$-values are assumed for the
  other two panels. The middle panel illustrates $T_{\rm osc}$ as a function of $m_a$, and the
  right panel shows the comoving wave number corresponding to the horizon at $T=T_{\rm osc}$.
  Shaded areas indicate constraints on the axion mass from the Lyman-$\alpha$ forest
  \citep{Haehnelt2017, Kobayashi2017}, solitonic core formation in Eridanus~II \citep{Marsh2018},
  and the spin-down of super-massive black holes via superradiant instability \citep{Stott2018}.
  The latter constraint is only shown in the left panel as these bounds disappear for
  $f_{a}\lesssim 10^{16}$~GeV due to ALP self-interactions \citep{Arvanitaki2015}.}
  \label{fig:ALP}
\end{figure}
\end{center}

Requiring that ALPs provide all of the DM fixes $f_{a}$ for given values of
$m_a$. This is shown in the left panel of \autoref{fig:ALP} for various choices
of $n$ and $b=1$. Using the approximate formula for the energy density,
\autoref{Eq:EnDenRes}, we can qualitatively estimate the behaviour as
\begin{equation}
f_a\propto b^{2/(8+3n)}m^{-(2+n)/(8+3n)}_{a}\;,
\label{Eq:Fproportionality}
\end{equation}
where we ignore a mild dependence of the proportionality constant on
$n$. Considering the large- and small-$n$ limit we find $f_{a}\propto
m^{-1/4}_a$ for $n=0$, and $f_{a}\propto m^{-1/3}_{a}$ for $n\gg 1$. In the
middle panel of the figure, we show $T_{\rm osc}$ as a function of $m_{a}$,
adopting values of $f_a$ that yield the correct DM abundance as shown
in the left panel. Qualitatively, we find $T_{\mathrm{osc}}\propto
m^{1/2}_a$ for $n=0$, and $T_{\mathrm{osc}}\propto m^{1/3}_a$ for $n\gg
1$. For the relevant region of parameter space, we see that $T_{\rm osc}$
is always much larger than the temperature at matter-radiation equality,
$T_{\rm eq} \sim$~few~eV. Hence, the ALP fields becomes matter-like early
enough to explain the DM in the Universe.

\subsection{Initial isocurvature power spectrum}
\label{sec:axions_power}

The random initial ALP field values in different Hubble patches will
lead to large variations in the energy density. Qualitatively, we
expect order-one density fluctuations between regions of the size of
the horizon at the time when ALP DM is born, i.e. at $T=T_{\rm osc}$
\citep{Hogan1988}. Densities in causally disconnected regions at that time
will be statistically uncorrelated. Let us denote the comoving wave
number corresponding to the horizon at $T=T_{\rm osc}$ by
\begin{equation}\label{eq:K}
K = R_{\rm osc} H(T_{\rm osc})\;.
\end{equation}
We then expect a constant power spectrum (white noise) of the ALP energy
density for $k < K$, with a cutoff around $k \simeq K$. Hence, the dimensionless
power spectrum, $\Delta^2$, can be parametrized as
\begin{equation}\label{eq:Delta}
  \Delta^2(k) \equiv \frac{k^3}{2\pi^2} P(k) = C
  \left(\frac{k}{K}\right)^3 \qquad (k \lesssim K)\;.
\end{equation}
For $C$ of $\mathcal{O}(1)$, the variance of the density fluctuations at scales
comparable to the horizon at $T_{\rm osc}$ is also of $\mathcal{O}(1)$. The shape
of the power spectrum for $k \sim K$ will be complicated and depends on the details
of the dynamics at temperatures around $T_{\rm osc}$. In the following, however, we
will be interested in length scales much larger than the horizon at $T_{\rm osc}$,
i.e. $k \ll K$. Thus, we do not need to know the precise shape around the cutoff, and
the parametrization \autoref{eq:Delta} will be an excellent approximation for the
scales of interest. Typical values of $K$ are shown in the right panel of \autoref{fig:ALP}.
During radiation domination, we approximately have $R\propto T^{-1}$ and $H\propto T^2$,
and using \autoref{eq:K} yields $K \propto T_{\mathrm{osc}}$. Therefore, $K$ and
$T_{\mathrm{osc}}$ have the same dependence on $m_a$, as is transparent from
\autoref{fig:ALP}.

To estimate $C$, the normalization constant of the power spectrum, we proceed as follows.
Using the result of \cite{Enander2017}, the energy density power spectrum is obtained
in terms of the ALP field's initial power spectrum as
\begin{equation}\label{eq:Pana}
  P(k) = 2(2\pi)^3\dfrac{\int d^3k^{\prime}\, P_\theta(\lvert\mathbf{k}^{\prime}\rvert )P_\theta (
  \lvert\mathbf{k} - \mathbf{k}^{\prime}\rvert )F(\mathbf{k}^{\prime}, \mathbf{k}^{\prime} - \mathbf{k})^2}
  {\left\lbrack\int d^3k^{\prime}\, P_\theta(k^{\prime}) F(\mathbf{k}^{\prime}, \mathbf{k}^{\prime}) \right\rbrack^{2}}\;,
\end{equation}
with $F(\mathbf{k}, \mathbf{k}^{\prime} )$ given by \autoref{eq:kernel}. Solving the ALP field's equations of motion to
determine $f_\mathbf{k}(t)$, we find that the power spectrum calculated according to \autoref{eq:Pana} becomes constant
in time shortly after $T_{\rm osc}$, as soon as the mass term dominates over the $k$-term in \autoref{eq:kernel}. Writing
\autoref{eq:Delta} as $P(k) = 2\pi^{2}C/K^{3}$, we obtain $C$ from the limit $k\to 0$ in \autoref{eq:Pana}. Depending on
the parameters $m_a$, $f_a$, $n$, and $b$, we arrive at values $0.04 \lesssim C \lesssim 0.3$.

The important observation is that these density fluctuations are of isocurvature type since they arise only in the DM
fluid. We will use that these fluctuations are also present at length scales relevant to CMB observations. As is seen
from the right panel of \autoref{fig:ALP}, the spectrum's cutoff scale is much larger than scales probed by the CMB,
$k_{\rm CMB}\ll K$. Since we are far away from the cutoff, the parametrization \autoref{eq:Delta} will give an accurate
description of the isocurvature power spectrum. For given values of $m_a$, $n$, and $b$ ($f_a$ fixed by assuming that
ALPs constitute all DM), we can predict the amplitude of the white noise power spectrum as outlined above, and test
whether CMB observations are compatible with the presence of such an isocurvature component.

\subsection{Assumptions and uncertainties}
\label{sec:axions_topo}

\paragraph{Harmonic approximation of the ALP potential.}

The assumptions and limitations of our approach are discussed in \cite{Enander2017}.
Let us briefly comment on the main uncertainties of our calculations and compare them
to other studies in the literature. An important assumption is the harmonic approximation
of the ALP potential which removes the periodic nature of the ALP field. Therefore, we
neglect the formation of cosmic strings and domain walls as well as contributions to the
ALP energy density from the decay of these topological defects shortly after the field
starts oscillating.\footnote{Note, however, that the existence of topological defects
during the early field evolution is a crucial ingredient for producing large fluctuations
in the axion field~\citep{Kibble1976}.}
These phenomena have been studied in some detail in the context of the QCD axion \citep[see,
e.g.,][]{Hiramatsu2012, Kawasaki:2014sqa, Klaer2017, Vaquero2018, Gorghetto2018, Kawasaki2018}.
Although the contribution of topological defects to the energy density can be substantial,
there is no general consensus on its quantitative size. This introduces an uncertainty of
$\mathcal{O}(1)$ in our estimate of the ALP energy density.

Further, the nonlinear structure of the field may also play an important role in the power
spectrum of density fluctuations. This has been recently studied in detail by numerical
simulations including the full periodic axion potential \citep{Vaquero2018}. Qualitatively,
the picture obtained in this work agrees with the results of \cite{Enander2017} for $k\ll K$.
In particular, it confirms the white noise power spectrum and supports our parametrization
\autoref{eq:Delta}, including the extrapolation to small $k$-values. Quantitatively, there
are some differences regarding the value of the coefficient $C$ appearing in \autoref{eq:Delta}.
Choosing values for $m_{a}$, $n$, and $b$ that correspond to the QCD axion, our method yields
$C\approx 0.15-0.16$ whereas the analysis of \citep{Vaquero2018} obtains $C = 0.03 \pm 0.01$,
approximately a factor 5 smaller than our semi-analytical result. We will use this as an estimate
for the systematic uncertainty in our prediction for the amplitude of the isocurvature power spectrum.
Below we will show results for $C$-values ranging from our estimate based on \autoref{eq:Pana} to
values that are 5 times smaller.

Note that the model dependence of $m_{a}(T)$ parametrized by $b$ and $n$ introduces an uncertainty
of similar size or larger. Also, our choices of the cutoff $Q$ in the initial power spectrum $P_\theta(k)$
as well as of the initial time ($T_{i} = 3T_{\rm osc}$) are somewhat arbitrary and introduce further
numerical uncertainties on the value of $C$ (see \cite{Enander2017} for a detailed discussion and some
quantitative estimates).
To summarize, although our method to compute the ALP energy density and the power spectrum is clearly
approximate, it leads to an order-of-magnitude estimate consistent with numerical simulations. Given the
even larger uncertainty due to the ALP model dependence, we thus proceed with our estimates.

\paragraph{Post-inflationary assumption.}

In our work, we always assume that the ALP field takes random values in causally disconnected regions
at temperatures well above $T_{\rm osc}$. This implies that the PQ symmetry is broken after the end of
inflation or restored at some point thereafter. The condition for this so-called \emph{post-inflationary
scenario} is \citep{Hertzberg2008}
\begin{equation}\label{eq:post-infl}
  f_{a} < {\rm max}\left\lbrack T_{\rm GH},\; T_{\rm max}\right\rbrack.
\end{equation}
Here the Gibbons-Hawkings temperature is defined by $T_{\rm GH} = H_{I}/2\pi$, where $H_{I}$ is the Hubble
parameter during inflation, and $T_{\rm max} = \epsilon_{\rm eff}E_{I}$ is the maximal temperature after
inflation, with $E_I$ denoting the energy scale of inflation, and $\epsilon_{\rm eff}$ is a dimensionless
efficiency parameter with $0 <\epsilon_{\rm eff} < 1$. Using $H_I = \sqrt{8\pi/3} E_I^2/M_P$, we have
$T_{\rm max} = \epsilon_{\rm eff} (3\pi/2)^{1/4} \sqrt{T_{\rm GH} M_P}$, which can be larger than $T_{\rm GH}$
for $\epsilon_{\rm eff} \simeq 1$.

Under the assumption of slow-roll inflation driven by a single scalar field with canonical kinetic term, CMB
data sets an upper limit on $E_{I}$ through the non-detection of primordial tensor modes, $E_{I} < 1.7\times 10^{16}$~GeV
(95\%~CL)~\cite{Akrami2018a}, which implies $T_{\rm GH} < 1.1\times 10^{13}$~GeV. Adopting this constraint, the
left panel of \autoref{fig:ALP} indicates that the condition $f_{a}< T_{\rm max}$ can be met in the relevant parameter
region for $m_{a} \gtrsim 10^{-19}$~eV, assuming $\epsilon_{\rm eff} \gtrsim 0.1$. A less model-dependent bound
on the scale of inflation has been derived in \cite{Abbott1984} using large-scale isotropy: $H_{I} < 10^{-4}M_{P}$.
This leads to upper bounds on $T_{\rm GH}$ and $T_{\rm max}$ about one order of magnitude larger than the ones
quoted above.

In summary, the condition \autoref{eq:post-infl} for the post-inflationary scenario can be satisfied under
reasonable assumptions on inflation in the region of ALP parameters that are of interest to us. As this generally
requires high values of $E_{I}$, we expect relatively large tensor-to-scalar ratios in such scenarios, likely to
be in the observable range in the near future (with some inflationary model-dependence).

\section{Constraints from cosmic large-scale structure}
\label{sec:cosmology}

\subsection{Large-scale imprint of isocurvature fluctuations}
\label{sec:cosmo_iso}
Fluctuations in the ALP density will affect the initial conditions of structure formation and their subsequent evolution through gravitational instability.
In what follows, we assume linear theory and focus on scalar perturbations to describe their imprint on CMB anisotropies and the matter power spectrum. The
initial perturbations are set deep within the radiation era where all modes of interest are well outside the horizon. Adopting the comoving (total-matter)
gauge, the standard initial conditions for adiabatic modes are related to the primordial spatial curvature perturbation, $\mathcal{R}$, generated by inflation
through \citep[e.g.,][]{Liddle2000}
\begin{equation}
\delta_{\mathbf{k}} = \dfrac{4}{9}\dfrac{1 + (2/5)X_{\nu}}{1 + (4/15)X_{\nu}}\left (\dfrac{k}{RH}\right )^{2}\mathcal{R}_{\mathbf{k}}\;,
\quad X_{\nu} = \dfrac{\rho_{\nu}}{\rho_{\gamma} + \rho_{\nu}}\;,
\label{eq:adiabatic}
\end{equation}
where $\delta = \delta\rho/\rho$ is the total density contrast, $\rho_{\gamma}$ and $\rho_{\nu}$ are the energy
densities of photons and neutrinos, respectively, and $X_{\nu} = (7/8)(4/11)^{4/3}N_{\rm eff}\approx 0.69$ for
an effective number of neutrino species $N_{\rm eff} = 3.046$. Adiabatic perturbations in single fluid components
$i$ characterized by equation-of-state parameters $w_{i}$ satisfy
\begin{equation}
\dfrac{\delta_{i}^{\rm ad}}{1+w_{i}} = \dfrac{3}{4}\delta^{\rm ad}\qquad \text{(adiabatic mode)}\;.
\label{eq:adiabatic2}
\end{equation}
As usual, we assume that $\Delta^{2}_{\mathcal{R}}$ takes the form of a nearly scale-invariant
spectrum parametrized by
\begin{equation}
\Delta^{2}_{\mathcal{R}} = A_{\rm s}\left (\frac{k}{k_{\ast}}\right)^{n_{\rm s}-1},
\quad n_{\rm s}\approx 1\;,
\label{eq:Delta_ad}
\end{equation}
where the amplitude $A_{\rm s}$ is defined with respect to the pivot scale $k_{\ast} = 0.05$ Mpc$^{-1}$.

In addition to the adiabatic mode, the breaking of the PQ symmetry after inflation will induce isocurvature perturbations
in the ALP field (see \autoref{sec:axions_power}). These may be written in terms of an initial entropy perturbation,
$\mathcal{S}_{a}$, defined relative to the photon component,
\begin{equation}
\mathcal{S}_{a} = \dfrac{\delta_{a}}{1+w_{a}} - \dfrac{3}{4}\delta_{\gamma}
= \dfrac{\delta_{a}^{\rm iso}}{1+w_{a}} - \dfrac{3}{4}\delta_{\gamma}^{\rm iso}
\approx \delta_{a}^{\rm iso} - \dfrac{3}{4}\delta_{\gamma}^{\rm iso} \approx \delta_{a}^{\rm iso}\;,
\label{eq:entropy}
\end{equation}
where we again assumed radiation domination, and the last step follows from the general isocurvature condition
$\sum_i\delta\rho_i^{\rm iso} = 0$ \citep[e.g.,][]{Liddle2000}. In \autoref{eq:entropy}, we have also set $w_{a}\approx 0$,
i.e. for the purposes of this work, we will use that the evolution of perturbations in the ALP field can be approximated
by that of a cold DM component. Although changes in $w_{a}$ are important at very early times, the ALP field quickly
adopts the behaviour of pressureless matter as soon as $T\lesssim T_{\rm osc}$. If the oscillations commence sufficiently
deep within the radiation era, ALPs can effectively be treated as a cosmic matter fluid, with initial conditions set
by eqs. \eqref{eq:adiabatic} and \eqref{eq:entropy}. Even so, the ALP field will exhibit an effective sound speed
\citep{Hwang2009, Park2012},
\begin{equation}
c_{s}^{2} = \dfrac{k^{2}}{k^{2} + 4m_{a}^{2}R^{2}}\;,
\label{eq:effective}
\end{equation}
that introduces a corresponding Jeans scale, $k_{J}$, below which the evolution of ALP density perturbations significantly differs from standard cold DM. Therefore, we
must additionally require $k \ll k_{J}$. Considering scales relevant to CMB observations, it turns out that both of the above criteria are already well satisfied for
ALPs with $m_{a}\gtrsim 10^{-24}$ eV \citep{Park2012, Hlozek2015, Hlozek2018}. Since the lower bounds on $m_{a}$ implied by the energy scale of inflation lie above this
threshold (see \autoref{sec:axions_topo}), our approximate treatment is justified.
\begin{center}
\begin{figure}
\includegraphics[trim= 0 -8 0 0,width=0.48\textwidth]{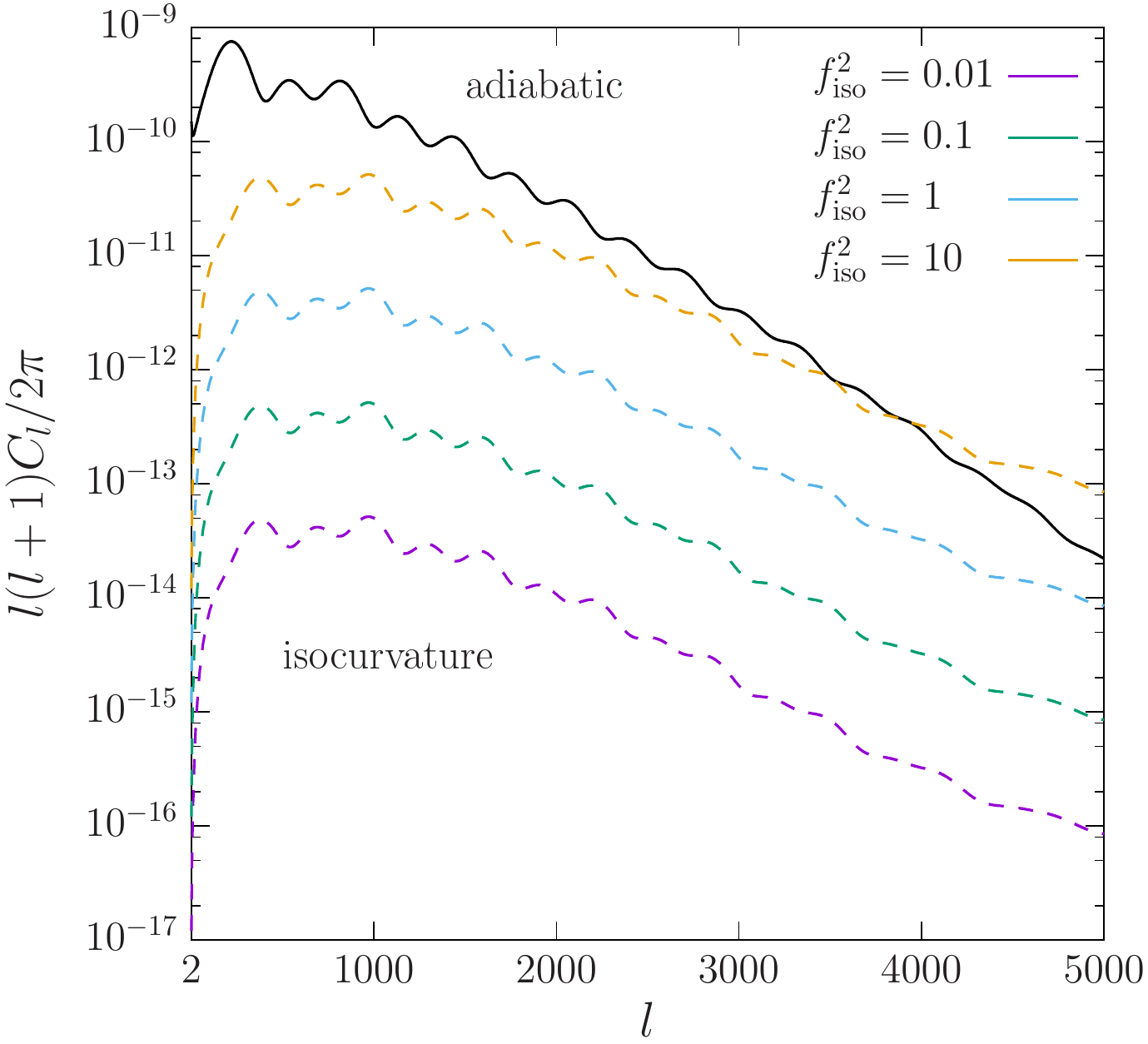}
\hfill
\includegraphics[width=0.48\textwidth]{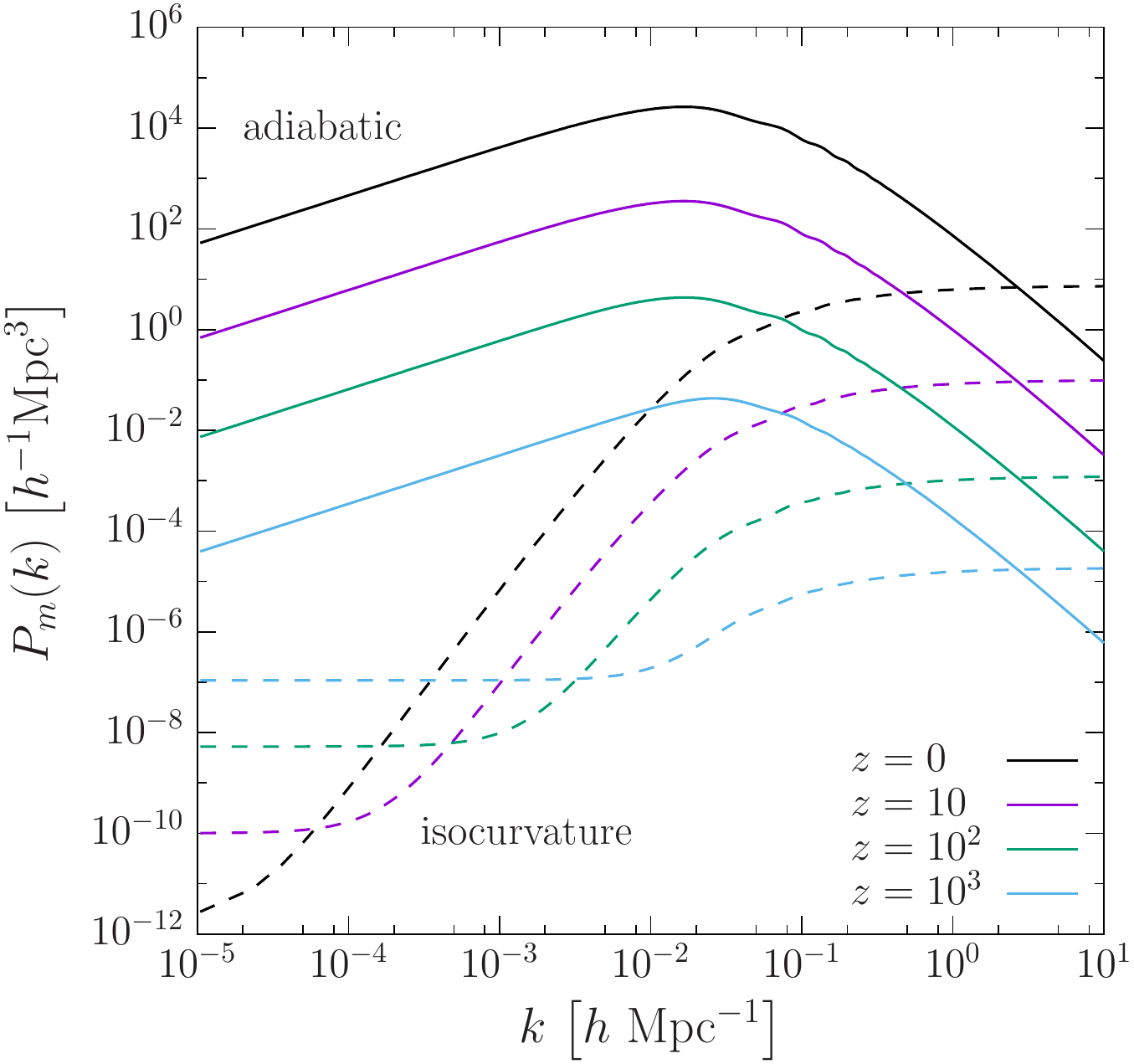}
\caption{Imprint of ALP DM isocurvature fluctuations (generated after inflation; dotted lines) on CMB anisotropies (unlensed; left) and the linear matter
power spectrum, $P_{m}(k)$, at various redshifts $0\leq z\leq 10^{3}$ for $f_{\rm iso}=0.1$ (expressed in comoving gauge; right). Results for the adiabatic
mode are shown as solid lines. As isocurvature and adiabatic modes are uncorrelated, total spectra are obtained by adding their individual contributions.}
\label{fig:iso_spectra}
\end{figure}
\end{center}
As discussed in \autoref{sec:axions_power}, the initial isocurvature spectrum, $\Delta^{2}_{\mathcal{S}}$, is then specified by \autoref{eq:Delta},
where the cutoff at $k\sim K$ can be safely ignored for practical purposes and is formally shifted to
infinity.\footnote{For the smallest masses considered here, this and the treatment of ALPs in terms of a cold DM component strictly hold on CMB scales
only. As we shall see shortly, however, current CMB data imply $m_{a}\gtrsim 10^{-20}$ eV, which allows us to use these assumptions on smaller scales
$k\lesssim 1$--$10h$ Mpc$^{-1}$ as well.}
Hence the ALP DM isocurvature mode is completely characterized by the amplitude of $\Delta^{2}_{\mathcal{S}}$ which is commonly parametrized in terms
of the entropy-to-curvature ratio, $f_{\rm iso}$, defined at the pivot scale,
\begin{equation}
f_{\rm iso}^{2} \equiv
\left. \frac{\Delta_{\mathcal{S}}^{2}}{\Delta_{\mathcal{R}}^{2}}\right\rvert_{k=k_{\ast}}\;.
\label{eq:fiso}
\end{equation}
In this work, we will derive constraints on $f_{\rm iso}$ based on Planck observations \citep{planck_collaboration_planck_2016, Planck2016_likelihoods} and adopt the
Fisher matrix formalism \citep{tegmark_karhunen-loeve_1997} to obtain forecasts for future CMB and 21-cm experiments. To solve the system of linearized Einstein and
fluid equations, we use the publicly available Boltzmann solver CLASS \citep{Lesgourgues2011}.

An example of how the ALP isocurvature perturbations generated after inflation affect CMB anisotropies and the linear matter power spectrum (in comoving gauge)
is depicted in \autoref{fig:iso_spectra}. The resulting spectra corresponding to adiabatic and isocurvature modes are shown as solid and dashed lines,
respectively. Since the two modes are assumed to be uncorrelated for ALP DM, the total spectra are given by the sum of the individual ones. Generally, the
imprint of isocurvature perturbations becomes more prominent with decreasing scale. For low-redshift observations such as galaxy surveys, nonlinearities
in the gravitational interaction are already important for $k\gtrsim 0.01h$ Mpc$^{-1}$, which complicates the interpretation of measurements and, most likely,
dilutes the ALP signal. However, probes of the matter power spectrum at higher redshifts (e.g., during the epoch of reionization), where gravitational nonlinearities
are much less developed, could provide interesting bounds on $f_{\rm iso}$ in addition to CMB observations.

\subsection{CMB experiments}
\label{sec:cmb_exp}

\begin{table}
\begin{center}
\begin{tabular}{cccc}
$\nu[\mathrm{GHz}]$ & $\theta_\mathrm{beam}[\mathrm{arcmin]}$ & $\sigma_T[\mu\mathrm{K}\;\mathrm{arcmin}]$ & $\sigma_P[\mu\mathrm{K}\;\mathrm{arcmin}]$ \\
\hline
\hline
90 & 5.7 & 18.80 & 26.6 \\
105 & 4.8 & 13.80 & 19.6 \\
135 & 3.8 & 9.85 & 13.9\\
160 & 3.2 & 7.78 & 11.0\\
185 & 2.8 & 7.05 & 9.97 \\
200 & 2.5 & 6.48 & 9.17\\
220 & 2.3 & 6.26  & 8.85 \\
\hline
\end{tabular}
\end{center}
\caption{Noise levels for a CMB stage IV experiment in different bands. The total noise contribution is the
inverse weighted sum of the individual noise contributions in each band.}
\label{tab:cmb}
\end{table}

\paragraph{Forecasts.}
The main CMB observables are the temperature ($T$) and polarization fluctuations ($E$- and $B$-mode polarization) which emerge from the
potential landscape and the anisotropy of Thomson scattering at the surface of last scattering, respectively. In what follows, we do not
consider secondary anisotropies such as the Sunyaev-Zeldovich effect \citep{sunyaev_microwave_1980} or CMB lensing \citep[e.g.,][]{hirata_reconstruction_2003}
and the integrated Sachs-Wolfe effect \citep{sachs_perturbations_1967}. Accordingly, we assume to have foreground-cleansed maps of the CMB
that can be decomposed into spherical harmonics with coefficients
\begin{equation}\label{eq:CMB_sperical_harmonics}
\hat a^X_{\ell m} = a^X_{\ell m} + n^X_{\ell m}\; ,
\end{equation}
where $a^X_{\ell m}$ denotes the signal, $n^X_{\ell m}$ the noise, and $X=T,E$ labels the corresponding probe. Here we ignore $B$-modes as
an additional probe since they only give rise to a tiny signal in the simplest inflationary scenarios and are primarily generated by CMB
lensing. For Gaussian random fields, all statistical properties are encoded in the two-point correlation function,
$C^{XY}_{\ell}=\langle a^X_{\ell m}a^{Y*}_{\ell m}\rangle$. To model the instrumental noise, we adopt the functional form
\begin{equation}
\label{eq:CMB_Noise}
N^{XX}_\ell = \sigma_X^2\exp\left\lbrack\ell(\ell+1)\frac{\theta^2_\mathrm{beam}}{8\log 2}\right\rbrack\;,
\end{equation}
where $\theta_\mathrm{beam}$ is the beam width and $\sigma_X^2$ is the noise of the measurement. The expected noise specifications for
CMB stage IV (s4) experiments \citep{abazajian_cmb-s4_2016} are summarized in \autoref{tab:cmb}.

Assuming Gaussian data and combining the information from all multipoles, the likelihood can be written as
\begin{equation}\label{eq:non-binned_likelihood}
p(\{\bm{a}_{\ell m}\}|\bm\theta) = \prod_\ell \left[\frac{1}{\sqrt{(2\pi)^2\det \bm C}} \exp\left(\bm a^\dagger_{\ell m} \bm {\bm C}^{-1} \bm a_{\ell m}\right) \right]^{2\ell +1}\;,
\end{equation}
where the $a^X_{\ell m}$ are assumed to be statistically independent.  To obtain forecasts, it suffices to adopt the
likelihood as given in \autoref{eq:non-binned_likelihood} from which the Fisher information matrix is readily obtained
as \citep{tegmark_karhunen-loeve_1997}
\begin{equation}\label{eq:Fisher_Gaussian}
F_{ij}(\bm\theta) = \sum_\ell\frac{2\ell +1}{2}\mathrm{tr}\left\lbrack\hat{\bm C}^{-1}\partial_{i}
\hat{\bm C} \hat{\bm C}^{-1}\partial_{j}\hat{\bm C}\right\rbrack_{\bm\theta}\; ,
\end{equation}
where hats indicate the inclusion of noise according to \autoref{eq:CMB_sperical_harmonics}, and $\partial_i$ denotes
the derivative with respect to the $i$-th parameter $\bm \theta_i$. Partial sky coverage of the data is taken into
account by multiplying \autoref{eq:Fisher_Gaussian} with the sky fraction $f_\mathrm{sky}$.
For CMB stage III (s3) experiments, we collect multipoles within the range $\ell=30$--2500, which is extended to
$\ell = 5000$ in both polarization and temperature for a CMB s4 survey. In both cases, the sky fraction is assumed
as $f_\mathrm{sky} = 0.7$.

\paragraph{Planck data.}
Additionally, we fit the above model to the 2015 Planck data release \citep{Planck2016_likelihoods} for which the likelihoods are
publicly available. Compared to the latest data release \citep{PLANCK2018parameters}, we note only slight changes in the overall
constraints on cosmological parameters. We use the \textit{Planck\textunderscore lite} likelihood which fits the $TT$ power spectrum
in the multipole range $\ell = 30$--2508. \textit{Planck\textunderscore lite} is a pre-marginalized version of the $TT$ likelihood
where all nuisance parameters (modulo Planck's absolute calibration) have been marginalized over prior to sampling. This choice
speeds up the analysis dramatically since the number of nuisance parameters is reduced to just one. To sample from the likelihood,
we use \textsc{MontePython} \citep{Montepython2013} which is interfaced with CLASS and sampling techniques such as \textsc{MultiNest}
\citep{skilling2006,2011Multinest} and \textsc{CosmoHammer} \citep{cosmohammer2013} embedding \textsc{EMCEE} \citep{goodman_ensemble_2010,Foreman-Mackey2013}.

\subsection{HI intensity mapping}
\label{sec:21cm_exp}
Next-generation radio telescopes such as the SKA \citep{SKA2015, SKA2018a, SKA2018b} are capable of probing fluctuations in the brightness temperature of the
21-cm neutral hydrogen line during the epoch of reionization with high sensitivity. At redshifts well before reionization is complete, the power spectrum of
these fluctuations should provide a close tracer of the underlying matter power spectrum and its measurement offers cosmological constraints that are
complementary to CMB and galaxy clustering observations. Reliably extracting the cosmological information from the 21-cm signal will be challenging due to
foreground contaminants and other astrophysical complexities, and it is presently unknown how well these systematics will be under control \citep{Furlanetto2006, Pritchard2012}.
To estimate future 21-cm constraints on a possible ALP isocurvature mode, we focus on statistical uncertainties and consider an optimistic scenario where density
fluctuations dominate the observed signal and all relevant astrophysical effects (including biases of the density field) can be accurately modeled or removed.

Assuming an average neutral hydrogen fraction $\overline{x}_{H}\approx 1$ and a spin temperature $T_{S}\gg T_{\rm CMB}$, fluctuations in the 21-cm brightness
temperature relative to the CMB at position $\mathbf{r}$ can be expressed as \citep[e.g.,][]{Zaldarriaga2004, McQuinn2006}
\begin{equation}
\delta_{21}(\mathbf{r}) = \frac{\Delta T_{21}(\mathbf{r})}{T_{0}} \approx \overline{x}_{H}\delta (\mathbf{r})\;,
\label{eq:temp21cm}
\end{equation}
where effects due to peculiar velocities have been neglected, $\delta$ is the matter density contrast, and $T_{0}$ denotes the average brightness temperature
for $\overline{x}_{H}= 1$ and an observed (redshifted) frequency $\nu =R\nu_{0}\approx 1420/(1+z)$ MHz \citep{McQuinn2006},
\begin{equation}
T_{0} \approx 26\dfrac{\Omega_{b}h^{2}}{0.022}\left (\dfrac{0.15}{\Omega_{m}h^{2}}\dfrac{1+z}{10}\right )^{1/2}\text{mK}\;.
\label{eq:t021cm}
\end{equation}
Using \autoref{eq:temp21cm}, the corresponding power spectra are simply related by $P_{21}\approx\overline{x}_{H}^{2}P_{\delta\delta}$. The expected error on
measurements of $P_{21}(k,\mu)$ for a single field at frequency $\nu = c/\lambda$ is given by a combination of cosmic variance and instrumental noise
\citep{Furlanetto2007, Lidz2008},
\begin{equation}
\sigma_{21}^{2}(k,\mu) = \left\lbrack P_{21}(k,\mu) + \frac{T_{\rm sys}^{2}}{T_{0}^{2}}\dfrac{D^{2}\Delta D}{Bt_{\rm int}n(k_{\bot})}
\left (\dfrac{\lambda^{2}}{A_{e}}\right )^{2}\right\rbrack^{2},
\label{eq:err1_temp21cm}
\end{equation}
where $\mu$ is defined such that $k_{\parallel}=\mu k$ and $k^{2}=k_{\bot}^{2}+k_{\parallel}^{2}$ (only modes in the upper half-plane are considered).
Here $T_{\rm sys}$ describes the system temperature, $D(z)$ is the comoving distance to the survey volume, $\Delta D$ denotes the survey depth, $B$ is
the bandwidth, and $t_{\rm int}$ is the total observation time. The effective collecting area per antenna tile, $A_{e}$, and the baseline density,
$n(k_{\bot})$, depend on instrumentation and array design. The error on the spherically averaged power spectrum can be obtained from
\begin{equation}
\dfrac{1}{\sigma_{21}^{2}(k)} = \sum_{\mu}\dfrac{k^{2}\Delta kV_{s}}{4\pi^{2}}\dfrac{\Delta\mu}{\sigma_{21}^{2}(k,\mu)}\;,\quad \mu > 0\;,
\label{eq:err2_temp21cm}
\end{equation}
where we choose a bin width $\Delta k=0.5k$ and $V_{s} = D^{2}\Delta D\lambda^{2}/A_{e}$ is the effective survey volume. For our calculations, we approximate
the sum in \autoref{eq:err2_temp21cm} by an integral.

Assuming that $T_{\rm sys}$ is set by the sky temperature, we take $T_{\rm sys}\approx 280\lbrack (1+z)/7.5\rbrack^{2.3}$ K \citep{Wyithe2007}. In addition,
we fix $B=8$ MHz to ensure a sufficiently small signal variation over the corresponding redshift range. The depth $\Delta D$ depends on $B$ and may be estimated
from
\begin{equation}
\Delta D \approx \dfrac{c\left (1+z\right )^{2}B}{\nu_{0}H(z)}\approx 1.7\left (\dfrac{B}{0.1\text{ MHz}}\right )\left (\dfrac{1+z}{10}\right )^{1/2}
\left (\dfrac{\Omega_{m}h^{2}}{0.15}\right )^{-1/2}\text{Mpc}\;.
\end{equation}
For an SKA-like survey, we further adopt $A_{e}\approx 290\text{ min}\lbrack\lambda^{2}/3, 3.2\text{ m}^{2}\rbrack$, yielding
$A_e(z=8)\approx 350\text{ m}^{2}$ and $A_e(z\gtrsim 14)\approx 925\text{ m}^{2}$ \citep{Pritchard2015}. In the continuous approximation, $n(k_{\bot})$
follows from a convolution of the antenna configuration. For simplicity, we assume a circularly symmetric array with constant baseline density up to a
maximum baseline $r_{\rm max}$,
\begin{equation}
n(\lvert\mathbf{u}\rvert ) = \dfrac{\lambda^{2}N_{a}(N_{a}-1)}{\pi r_{\rm max}^{2}}
\approx\dfrac{\lambda^{2}N_{a}^{2}}{\pi r_{\rm max}^{2}}\;,\quad \lambda\lvert\mathbf{u}\rvert < r_{\rm max}\;,
\end{equation}
and zero otherwise, where $\mathbf{u}$ is the dimensionless baseline vector in visibility space, satisfying $\lvert\mathbf{u}\rvert = k_{\bot}D(z)/2\pi$, and
$N_{a}$ is the total number of antennae. Setting $r_{\rm max} = 1$ km, we obtain an average sensitivity $\sim 0.9$, which is in good agreement with typically
adopted specifications \citep{Pritchard2015, Chang2015}. The discrete nature of the array configuration prohibits measurements for arbitrarily small baselines.
For an SKA-like survey, we choose a minimum baseline $r_{\rm min}= 35$ m.

Below we will consider two different stages of an SKA-like experiment. For an idealized SKA1 setup targeting redshifts $z\gtrsim 6$, we take $N_{a} = 900$, and
we assume that this number further increases to $N_{a} = 3600$ in an SKA2 phase \citep{Pritchard2015}. To account for foregrounds, we follow previous works
\cite[e.g.,][]{McQuinn2006, Lidz2008} by imposing a lower bound on the observable wavelengths. The cleaning process relies on the expectation that foregrounds
are spectrally smooth, whereas the signal has structure in frequency space. At the very least, this will remove all line-of-sight modes with
$k_{\parallel}\leq 2\pi/\Delta D$. From the discreteness of modes in the survey it then follows that all modes satisfying
\begin{equation}
k\leq 2\pi/\Delta D
\end{equation}
will be lost. Considering a fiducial survey volume at $z\sim 8$ and restricting the analysis to linear modes, significant constraints on the matter power
spectrum are typically obtained for $k\sim 0.1$--$1h$ Mpc$^{-1}$. Concerning the computation of the Fisher matrix, we assume measurements in seven non-overlapping
bins covering the range $k\approx 0.08$--$2.7h$ Mpc$^{-1}$, with $\Delta k$ as specified above. Since the CMB probes the power spectrum at much larger scales,
the corresponding likelihoods may simply be added to arrive at combined forecasts.

\section{Results}
\label{sec:results}

In this section, we discuss forecasts on constraints from CMB anisotropies and HI intensity mapping on the isocurvature
amplitude, as well as bounds based on current Planck data (\autoref{sec:constr}). These are then converted into constraints
on ALP masses (\autoref{sec:ALPbounds}). Regarding our analysis, we will adopt two cosmological models, one for the
Fisher forecasts specified by a total of ten parameters (listed in \autoref{tab:params}), and a standard flat six-parameter
cosmological model ($\Lambda$CDM) extended by $f_{\rm iso}$ and used to fit the Planck likelihood described in \autoref{sec:cmb_exp}.

\begin{table}
\begin{scriptsize}
\begin{center}
  \begin{tabular}{ccccccccccc}
Experiment & $f_\mathrm{iso}$ & $\alpha_\mathrm{s}$ & $\sum m_\nu[\mathrm{eV}]$ & $n_\mathrm{s}$ & $A_\mathrm{s}$ & $\Omega_\mathrm{b}$ & $\tau$ & $h$ & $\Omega_\mathrm{m}$ & $\bar{x}_H$\\
  \hline
  \hline
  s3 & 0.38 & 0.0052 & 0.34 & 0.0034 & 0.021 & 0.0045 & 0.0045 & 0.032 & 0.038 & -\\
  s3$+$SKA1 & 0.20 & 0.0044 & 0.28 & 0.0031 & 0.021 & 0.0037 & 0.0043 & 0.027 & 0.031 & 0.082 \\
  s4 & 0.067 & 0.0018 & 0.050 & 0.0016 & 0.0080 & 0.00064 & 0.0017 & 0.0045 & 0.0053 & - \\
  s4$+$SKA2 & 0.016 &0.0017 & 0.042 & 0.0016 & 0.0080 & 0.00051 & 0.0017 & 0.0034 & 0.0040 & 0.012 \\
\hline
\end{tabular}
\end{center}
\end{scriptsize}
\caption{Parameters used in the Fisher forecasts outlined in sections \ref{sec:cmb_exp} and \ref{sec:21cm_exp}. Shown
are the marginalized 1$\sigma$-errors of each parameter for the experiment(s) given in the first column. For both SKA1
and SKA2, we assume single-field observations at $z=8$ and $t_{\rm int}= 2000$ h. As described in the text, the reference
model adopts $\overline{x}_{H}=1$ and $f_{\rm iso}=0.005$, other fiducial values are summarized at the end of \autoref{sec:intro}.}
\label{tab:params}
\end{table}

\subsection{Bounds on the isocurvature amplitude}
\label{sec:constr}
Concerning forecasts based on the Fisher matrix formalism, we consider the set of parameters presented in \autoref{tab:params}.
In particular, these refer to the spectral index of the primordial power spectrum, $n_{\rm s}$, its running $\alpha_{\rm s}$
and amplitude $A_{\rm s}$ (defined at the pivot scale $k_{\ast} = 0.05$ Mpc$^{-1}$), the sum of neutrino masses $\sum m_{\nu}$,
the Hubble constant $H_{0} = 100h$ km s$^{-1}$ Mpc$^{-1}$ expressed in terms of the dimensionless Hubble parameter $h$, the
optical depth $\tau$, and the baryon and total matter density parameters $\Omega_{\rm b}$ and $\Omega_{\rm m}$, respectively.
Further, $f_{\rm iso}$ describes the relative amplitude of isocurvature fluctuations as defined in \autoref{eq:fiso}, and
$\overline{x}_{H}$ is the neutral hydrogen fraction that enters in the analysis of 21-cm experiments discussed in \autoref{sec:21cm_exp}.
Note that the ALP DM density parameter $\Omega_{\rm a}$ in presence of neutrinos is given by
\begin{equation}
\Omega_{\rm a} \equiv \Omega_{\rm m} - \Omega_{\rm b} - \frac{\sum m_{\nu}}{ 93.14 h^{2}}\;.
\end{equation}
For our reference model, we set $\overline{x}_{H}=1$ and $f_{\rm iso}=0.005$. The small, but non-zero value of $f_{\rm iso}$ is
chosen to avoid numerical issues in the computation of the Fisher matrix and should be understood as a proxy for a vanishing
isocurvature component. The fiducial values of all remaining cosmological parameters can be found at the end of \autoref{sec:intro}.

\begin{center}
\begin{figure}
\includegraphics[width=0.9\textwidth]{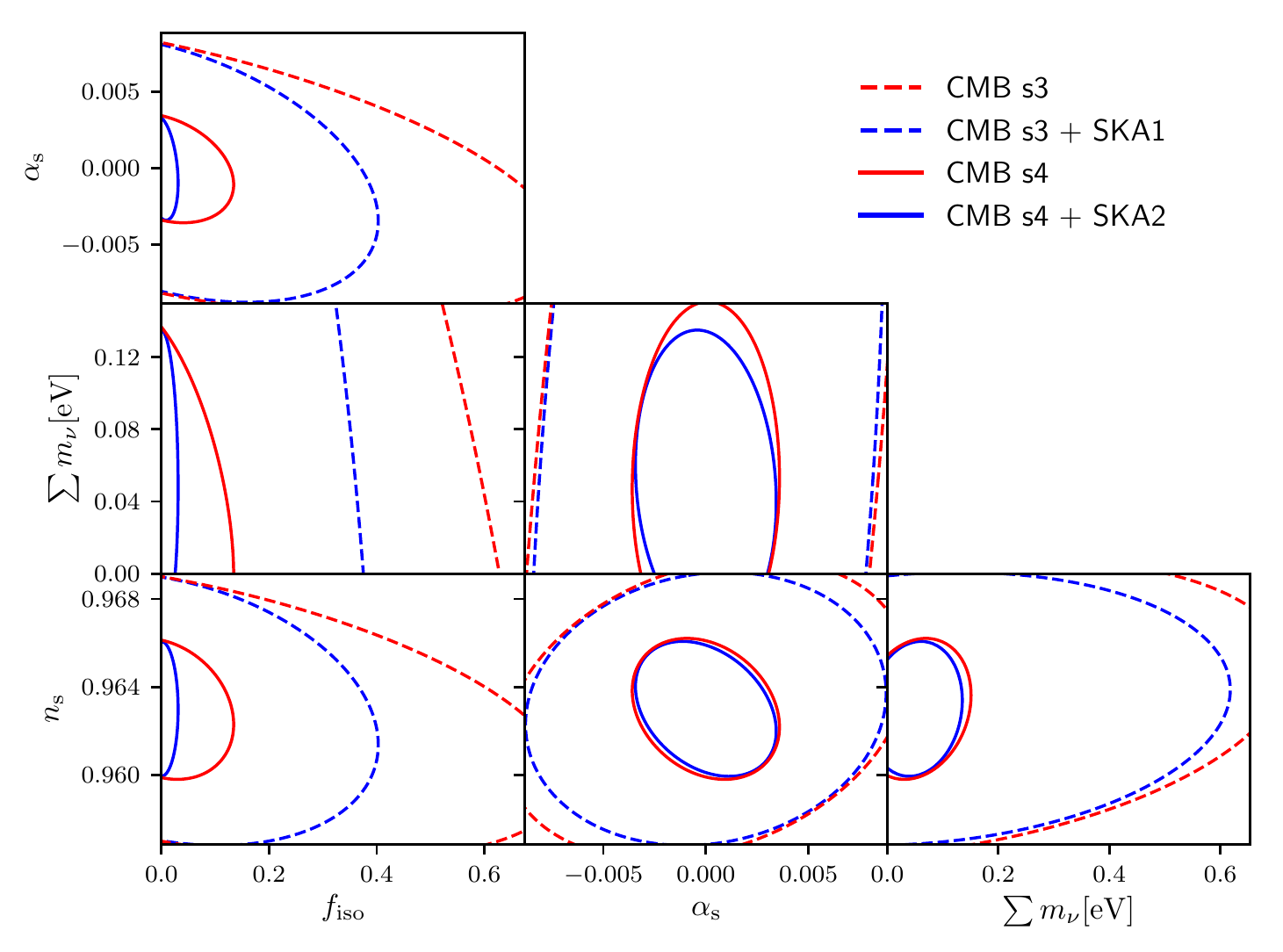}
\caption{Marginalized $1\sigma$-contours for the parameters $f_{\rm iso}$, $n_{\rm s}$, $\alpha_{\rm s}$, and $\sum m_{\nu}$
obtained from the Fisher analysis, assuming the parameters listed in \autoref{tab:params}. Dashed contours indicate results
for CMB s3 surveys while solid curves correspond to future s4 experiments. Red curves show constraints based on CMB observations
alone, blue curves refer to combinations with HI intensity mapping experiments. For both SKA1 and SKA2, we assume single-field
observations at $z=8$ and $t_{\rm int}= 2000$ h.}
\label{fig:fisher}
\end{figure}
\end{center}

Focusing on the parameters $f_{\rm iso}$, $n_{\rm s}$, $\alpha_{\rm s}$, and $\sum m_{\nu}$, \autoref{fig:fisher} shows the resulting marginalized 1$\sigma$-contours
for CMB s3 and s4 experiments as well as for their combination with the idealized 21-cm surveys from \autoref{sec:21cm_exp}. For both SKA1 and SKA2,
we assume single-field observations at $z=8$ and a total observation time $t_{\rm int}= 2000$ h, which roughly translates into a 2-year measurement
period. The corresponding $1\sigma$-constraints for all parameters and the same combinations of experiments are summarized in \autoref{tab:params}.
From the figure, we clearly see that the constraints on $f_\mathrm{iso}$ are greatly improved when considering a CMB s4 experiment alone. This is in
large part due to the drastically increased sensitivity of these experiments in polarization measurements. Also, the general trend of tighter bounds
from future surveys can be understood from their improved ability to probe higher wave numbers where the imprint of the ALP isocurvature mode becomes
more pronounced (see \autoref{sec:cosmo_iso}). As we will discuss in the context of Planck data below, the condition $f_{\rm iso}\geq 0$ hampers a
straightforward interpretation of the inferred $f_{\rm iso}$-errors from the Fisher analysis in terms of usual confidence intervals. Therefore, the
quoted uncertainties on $f_{\rm iso}$ should be taken as rough estimates.

Looking at \autoref{fig:fisher}, we further notice degeneracies between $f_\mathrm{iso}$ and other parameters that determine the shape of the matter
power spectrum. While the resulting degeneracy with the sum of neutrino masses appears to be very weak, it is found to be quite appreciable for
the two shape parameters $n_\mathrm{s}$ and $\alpha_\mathrm{s}$. Interestingly, combining CMB observations with HI intensity mapping experiments
can break these degeneracies to some extent. In general, we see that the addition of 21-cm observations significantly tightens the constraints on
$f_\mathrm{iso}$. Considering our estimates for the SKA2 survey, the bound is reduced by another factor of 5 compared to a pure CMB experiment,
resulting in the limit $f_\mathrm{iso} \leq 0.016$. Other constraints could come from even smaller scales or, for instance, from weak lensing analyses.
However, this would require a thorough understanding of the evolution of isocurvature fluctuations on scales that are already in the nonlinear
regime of structure formation.

\begin{figure}
\begin{center}
\includegraphics[width=0.475\textwidth]{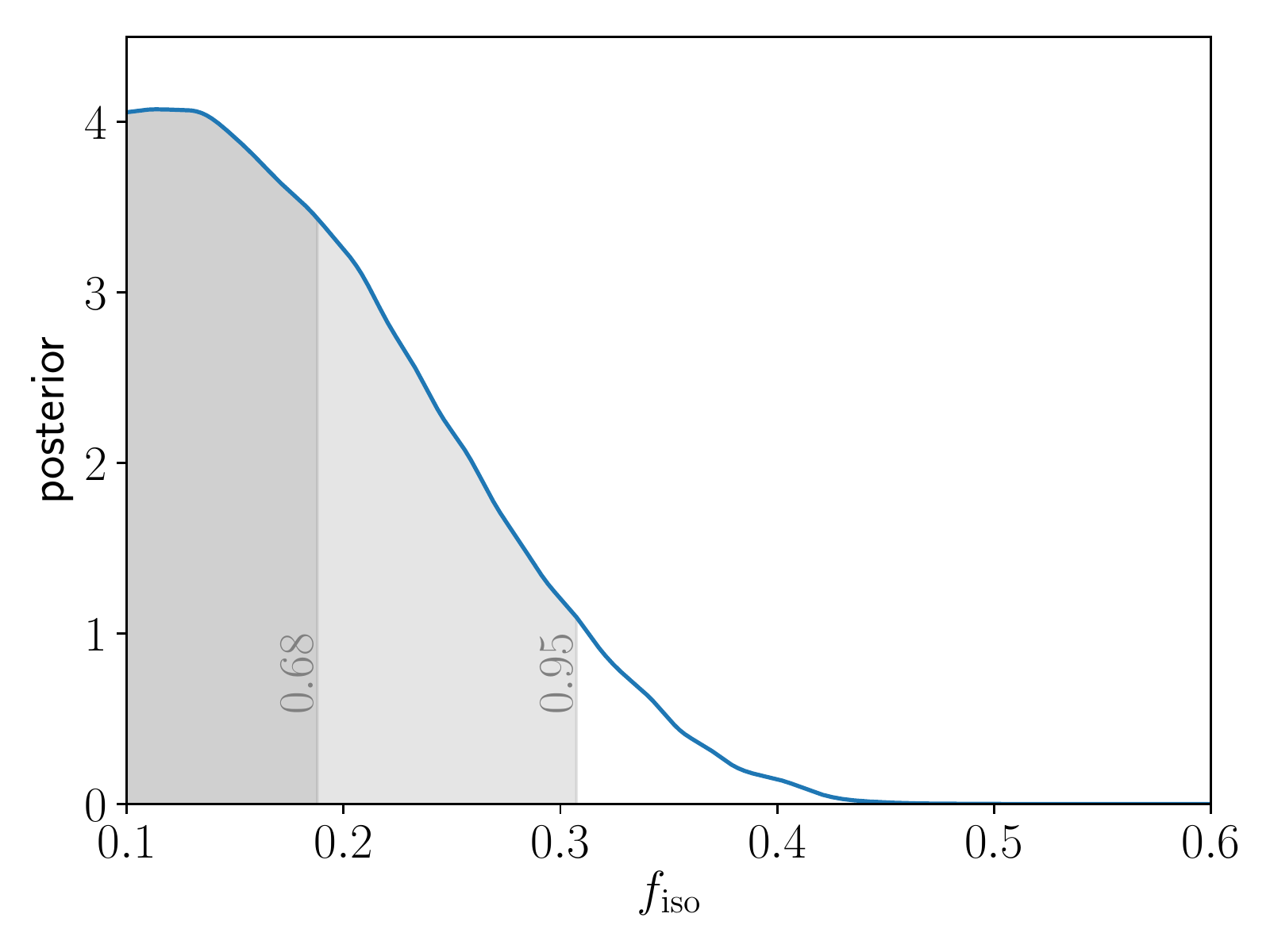}
\hfill
\includegraphics[width=0.475\textwidth]{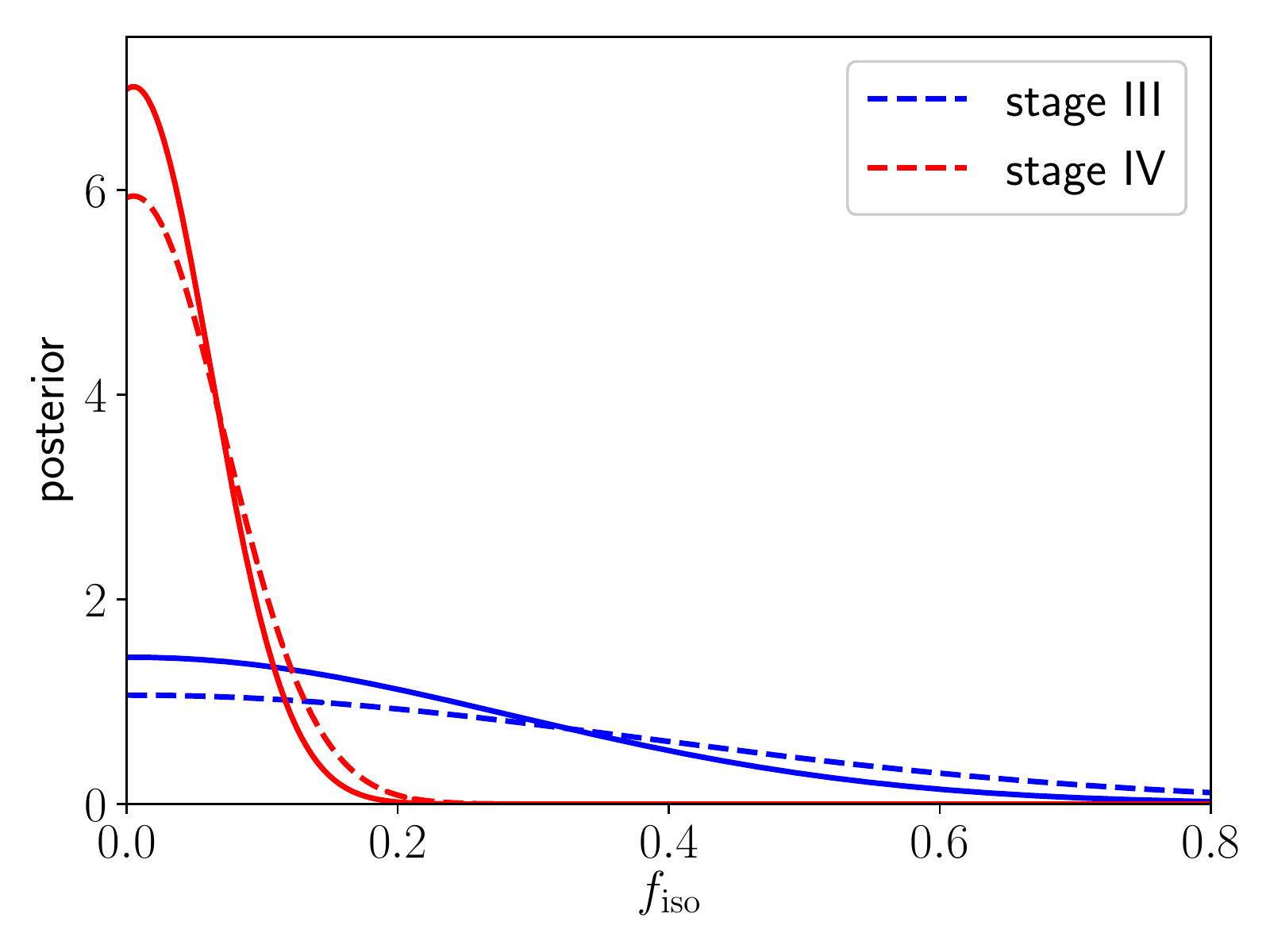}
\end{center}
\caption{
Posterior distribution of the isocurvature amplitude $f_\mathrm{iso}$. The left panel illustrates the posterior distribution
from \textit{Planck\textunderscore lite} including corresponding confidence regions. The right panel shows the effect
of marginalizing over $\alpha_{\rm s}$ and $\sum m_{\nu}$ on CMB forecasts. Dashed curves indicate the posterior after
marginalizing over the full cosmology as summarized in \autoref{tab:params}. Solid curves show the posterior when both
$\alpha_{\rm s}$ and $\sum m_{\nu}$ are kept fixed.}
\label{fig:PLANCK_compare}
\end{figure}

Finally, we study constraints on $f_\mathrm{iso}$ based on the Planck 2015 data. To this end, we fit a standard 6-parameter $\Lambda$CDM model
extended by $f_{\rm iso}$ to the temperature power spectrum as described in \autoref{sec:cmb_exp}. The results are presented in the left
panel of \autoref{fig:PLANCK_compare} where we plot the marginalized posterior distribution of $f_\mathrm{iso}$. It is found that Planck
yields a 2$\sigma$-bound of $f_\mathrm{iso}\leq 0.31$.
In the right panel of \autoref{fig:PLANCK_compare}, we illustrate the impact of reducing the model's parameter space in the Planck data
analysis. Omitting the spectral running, $\alpha_\mathrm{s}$, and the sum of the neutrino masses, $\sum m_\nu$, tightens the posterior
slightly and leads to a small improvement of the constraints. A few comments on these two plots are in order. The constraints from
Planck outperform the CMB s3 forecast even if $\alpha_\mathrm{s}$ and $\sum m_\nu$ are not marginalized over. The reason for this is
that we evaluate the Fisher matrix in a region where the curvature of the log-likelihood is relatively small since $f_\mathrm{iso}$
is bounded by zero. Therefore, the actual posterior distribution is very asymmetric and turns very flat towards zero. Hence, the
constraints derived from the Fisher matrix turn out weaker. Note that the addition of the $E$-mode spectrum in the forecasts, which is
not included in the \textit{Planck\textunderscore lite} likelihood, yields only small changes in the constraints due to the high noise
levels assumed for CMB s3 polarization measurements.

It is interesting to compare our results to the isocurvature bounds obtained by the Planck collaboration \citep{Planck2015inflation,Akrami2018a}.
Their analysis uses different models to constrain the isocurvature component. The most general one assumes a free power law for the isocurvature,
adiabatic and cross fluctuations. The power law is constructed between $k=0.002$--0.1 Mpc$^{-1}$, i.e. over scales accessible to Planck. For
these models, it is found that $\beta\leq 0.37$ at $k_* = 0.05\,\mathrm{Mpc}^{-1}$, where $\beta\equiv f_\mathrm{iso}^2/(1+f_\mathrm{iso}^2)$.
Our bound $f_{\rm iso} \leq 0.31$ corresponds to $\beta\leq 0.088$. Since the model presented here consists of fewer parameter (in particular,
the slope of the isocurvature power spectrum is fixed by the ALP model), these estimates appear consistent with each other. A specific axion model
with fixed primordial tilt (spectral index) $n_{\rm s}^{\rm iso} = 1$ and free amplitude is considered in \cite{Planck2015inflation}, leading to $\beta \lesssim 0.04$.
This model, however, requires PQ symmetry breaking before the end of inflation, in contrast to our assumption. Similarly, axion models with free
spectral tilt \citep{Akrami2018a} cannot be directly compared to our case (where $n_{\rm s}^{\rm iso} = 4$ is fixed).

\subsection{Implications for ALP masses}
\label{sec:ALPbounds}

Using our prediction for the ALP isocurvature power spectrum \autoref{eq:Delta} and the parametrization of the adiabatic spectrum in
\autoref{eq:Delta_ad}, we can relate the value of $f_{\rm iso}$ to the underlying ALP parameters,
\begin{equation}\label{eq:fiso-pred}
  f_{\rm iso} = \sqrt{\frac{C k_{\ast}^{3}}{A_{\rm s}K^{3}}}\;,
\end{equation}
where the values of $C$ and $K$ are calculated as described in \autoref{sec:axions}. Assuming the fiducial value
$A_{\rm s} = 2.215\times 10^{-9}$, we see that the isocurvature fraction on CMB scales becomes of order one for
$K/k_* \sim (A_{\rm s}/C)^{-1/3} \sim 1000$.

\begin{center}
\begin{figure}
\includegraphics[width=0.32\textwidth]{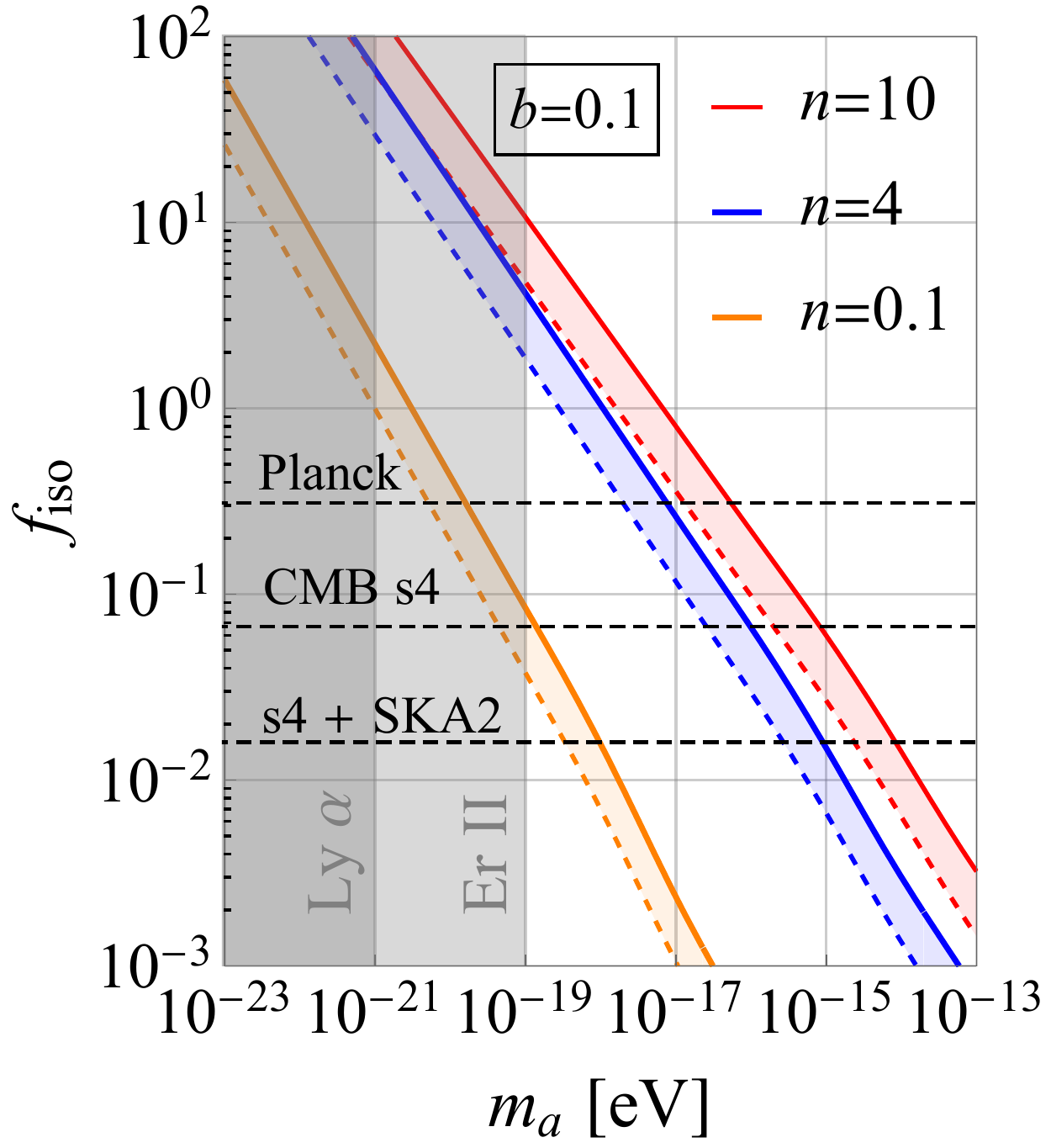}
\includegraphics[width=0.32\textwidth]{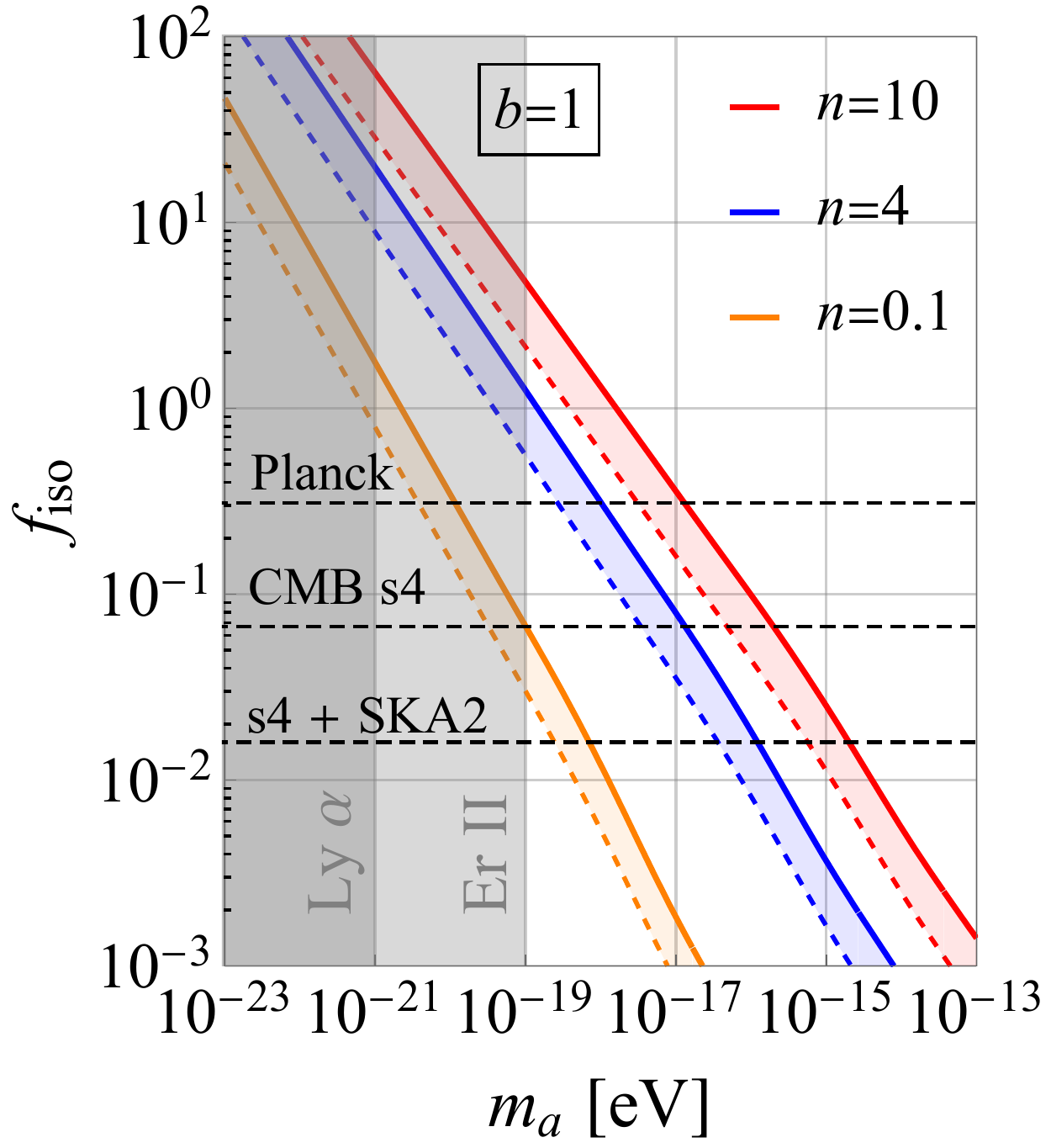}
\includegraphics[width=0.32\textwidth]{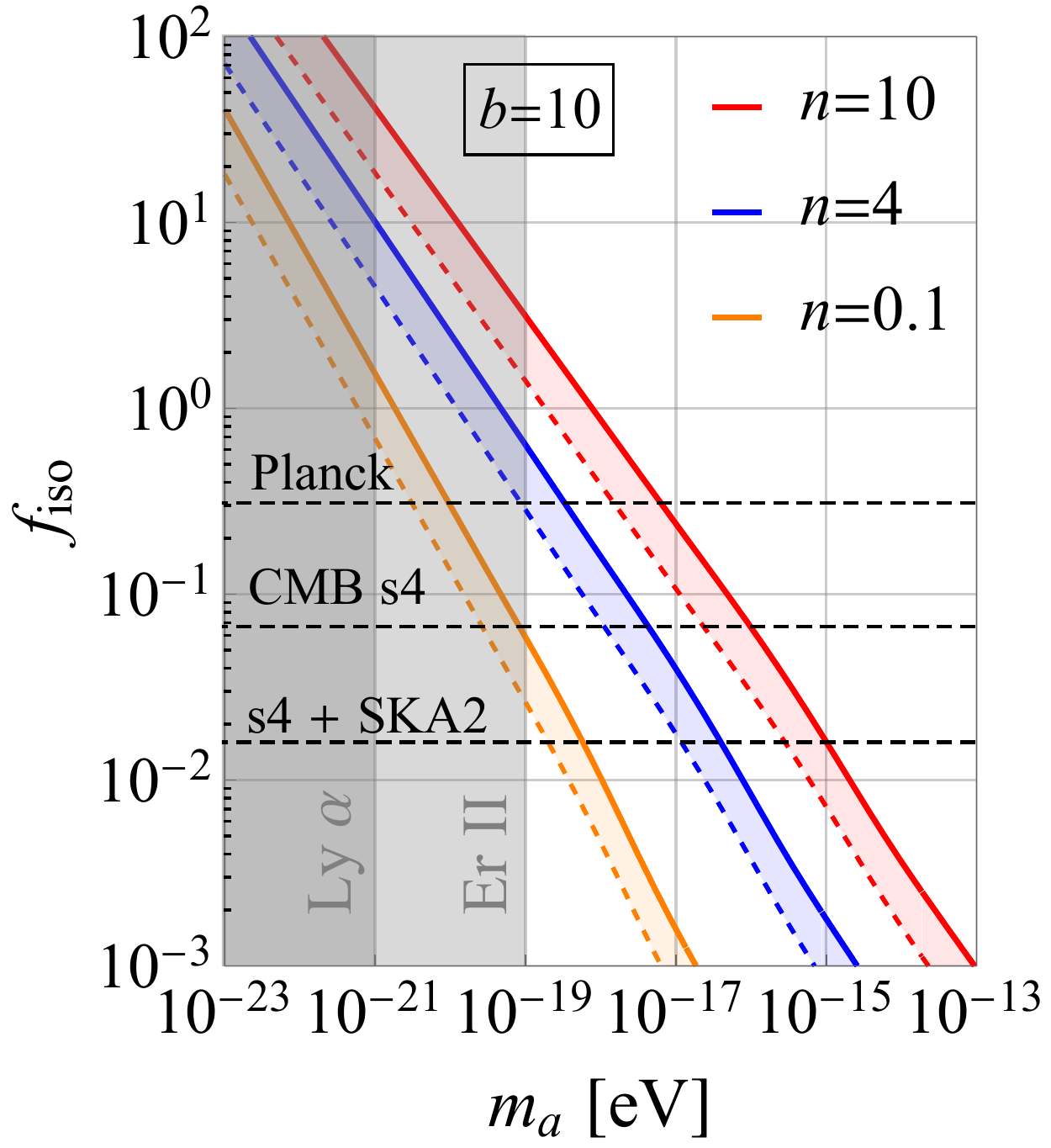}
\caption{Predicted isocurvature fraction according to \autoref{eq:fiso-pred} as a function of the zero-temperature ALP
  mass $m_a$ for different assumptions on the ALP mass-temperature dependence parametrized by $n$ and $b$ as defined in
  \autoref{Eq:MassParameterization}. Solid curves correspond to our estimate of the isocurvature amplitude, and bands
  between solid and dashed curves indicate a factor 5 uncertainty. Horizontal dotted lines show upper bounds on $f_{\rm iso}$
  from Planck ($2\sigma$-level) and the sensitivity of future CMB s4 and advanced 21-cm experiments (SKA2). Shaded regions
  correspond to those of \autoref{fig:ALP}. The PQ breaking scale $f_a$ is fixed for each $m_a$ by requiring that ALPs provide all DM.
  \label{fig:fiso-ma}}
\end{figure}
\end{center}

In \autoref{fig:fiso-ma}, we show the predicted values of $f_{\rm iso}$ as a function of the
ALP mass $m_a$ for different assumptions on the mass-temperature dependence $m_{a}(T)$. The
width of the bands indicates a factor 5 systematic uncertainty in predicting the amplitude,
i.e. the constant $C$ in \autoref{eq:fiso-pred}, as we discussed in \autoref{sec:axions_topo}.
The qualitative behaviour of these curves follows from the $K$-dependence shown in the right
panel of \autoref{fig:ALP} since, up to a minor dependence on $C$, we roughly have $f_{\rm iso}
\propto K^{-3/2}$ from \autoref{eq:fiso-pred}. From the estimates in \autoref{sec:axions_cosmo},
we, therefore, expect that
$f_{\rm iso} \propto m_a^{-3/4}$ ($m_a^{-1/2}$) in the
small-$n$ (large-$n$) limit, in good agreement with the figure.

\begin{center}
\begin{figure}
  \includegraphics[width=0.32\textwidth]{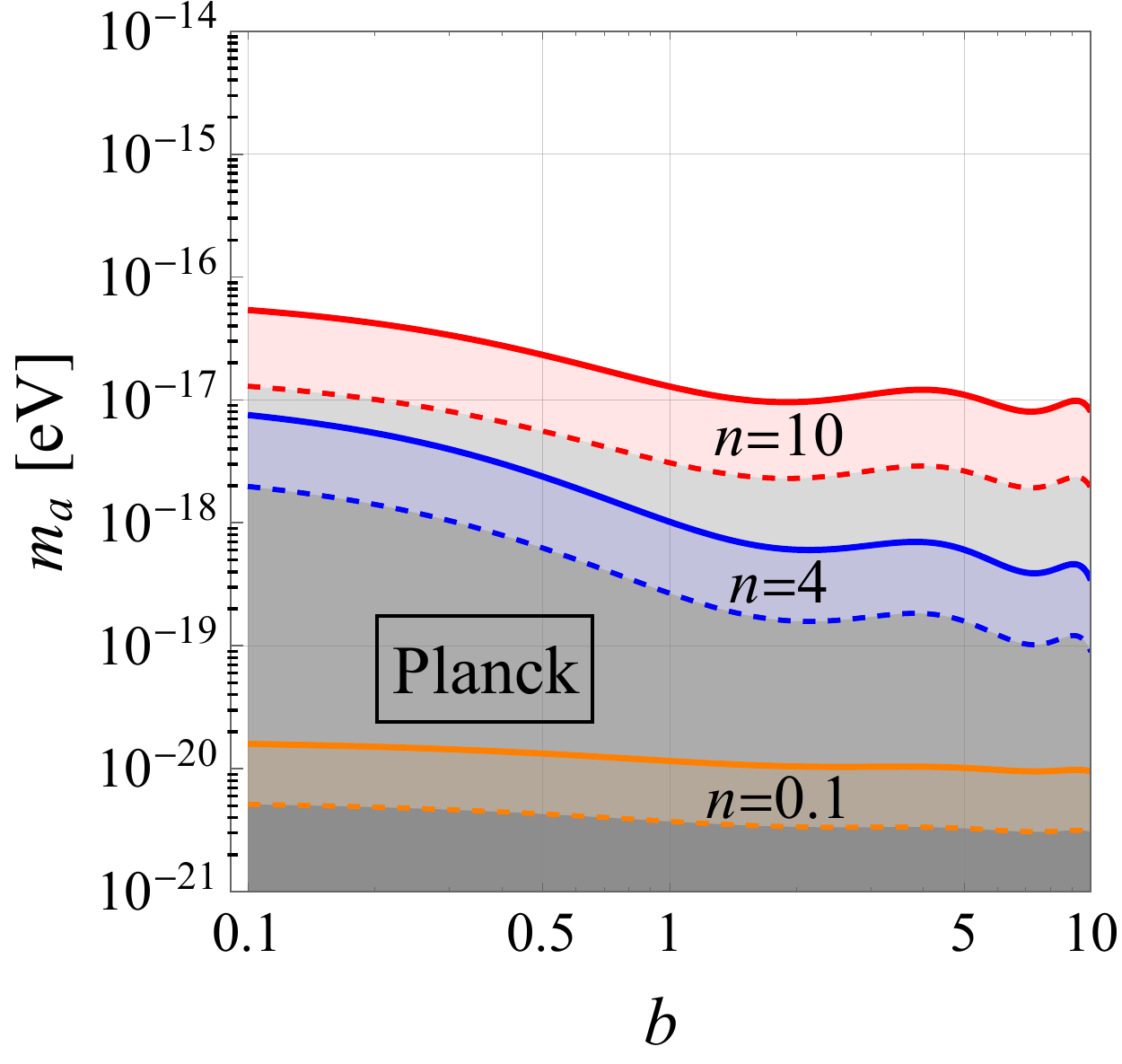}
  \includegraphics[width=0.32\textwidth]{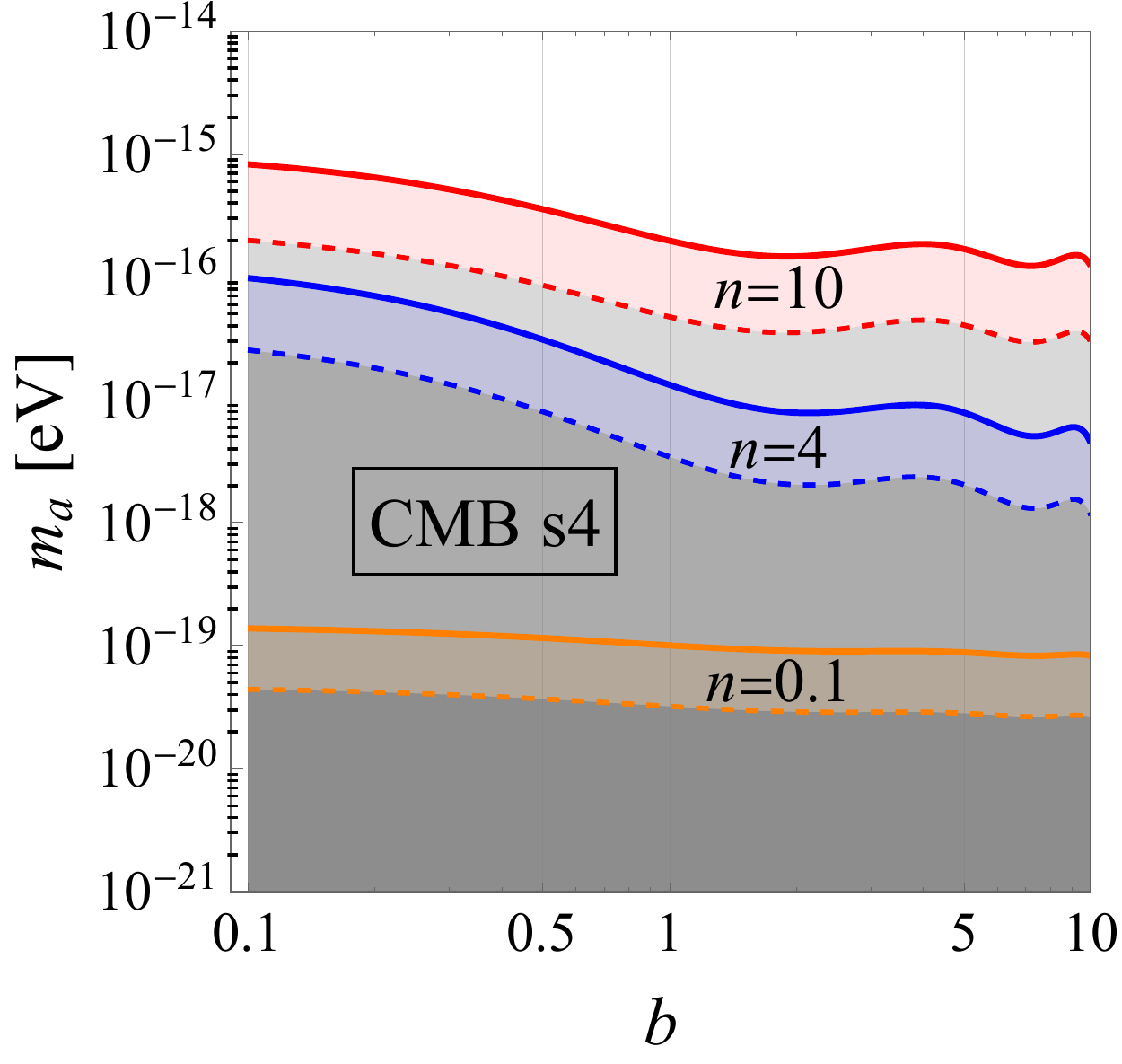}
  \includegraphics[width=0.32\textwidth]{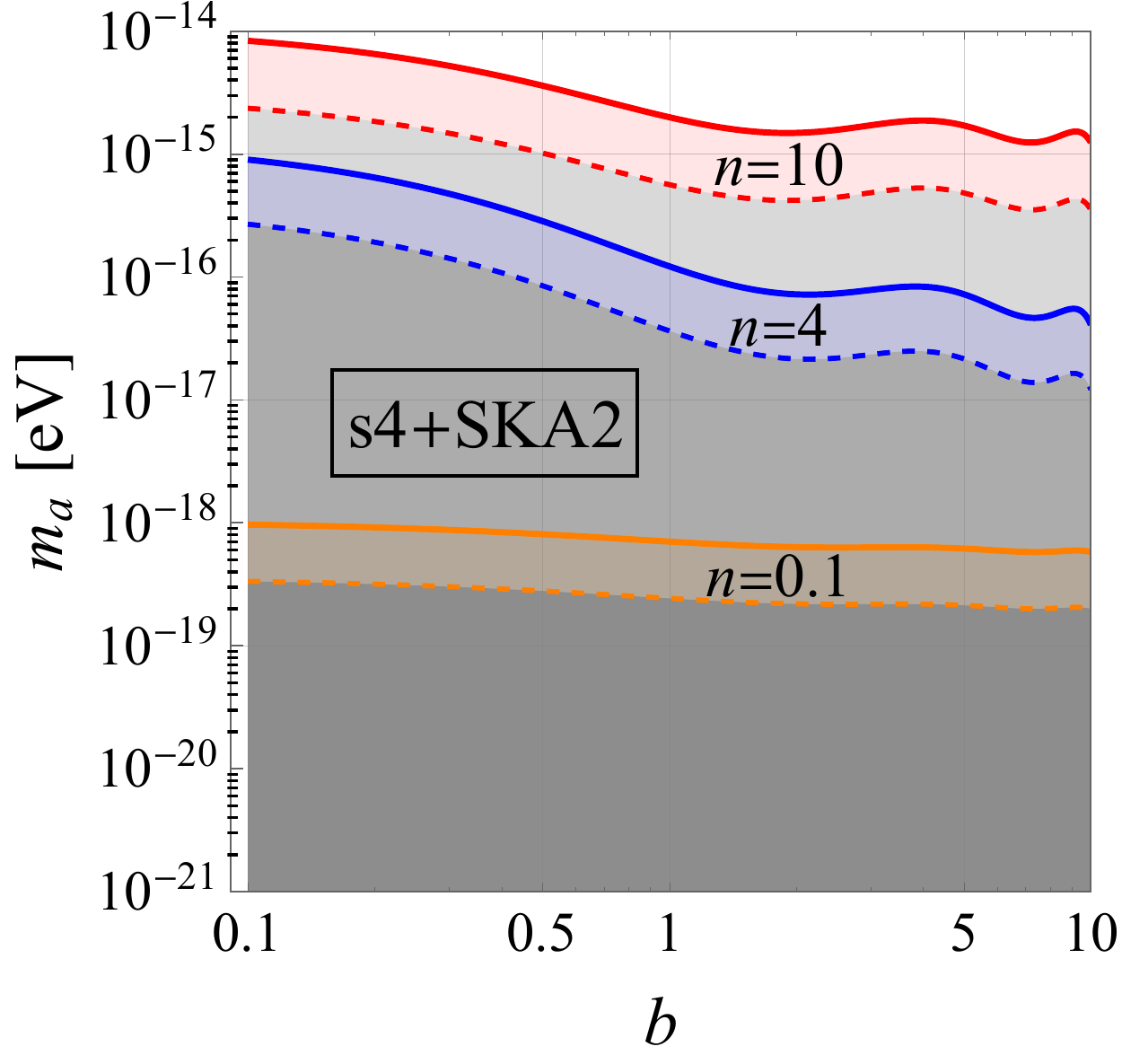}
  \caption{Constraints on the zero-temperature ALP mass $m_{a}$ from isocurvature fluctuations for different
  	assumptions on the mass-temperature dependence by $n$ and $b$ as defined in \autoref{Eq:MassParameterization}.
  	ALPs are assumed to provide all DM. Solid curves correspond to our estimate of the isocurvature amplitude,
  	and bands between solid and dashed curves indicate a factor 5 uncertainty. The left panel corresponds to
  	$f_{\rm iso} < 0.31$ ($2\sigma$-level) from Planck. The middle and right panels show, respectively, potential
  	bounds assuming $1\sigma$-sensitivities of a future CMB s4 experiment ($f_{\rm iso} < 0.067$) and its combination
  	with an SKA2-like 21-cm survey ($f_{\rm iso} < 0.016$). The region below the curves is disfavoured.}
\label{fig:ma-bound}
\end{figure}
\end{center}

For $m_{a}\lesssim 10^{-16}$~eV, we observe from \autoref{fig:fiso-ma} that the amplitude of isocurvature
fluctuations with $k \sim k_*$ can become comparable to the adiabatic modes and, therefore, relevant to
CMB observations. The predicted values of $f_{\rm iso}$ are compared to the CMB bounds implied by Planck,
as well as the sensitivity of future CMB s4 and 21-cm experiments. In the regions of parameter space where,
the predictions for $f_{\rm iso}$ exceed the Planck constraint $f_{\rm iso} < 0.31$, the assumption of
post-inflationary ALP DM is excluded at the $2\sigma$-level. This excluded region is illustrated in the
left panel of \autoref{fig:ma-bound}. The exclusion is stronger for large values of $n$ and small values
of $b$. For $n = 10$, we see that values of $m_a \lesssim 10^{-17}$~eV are excluded. Even in the small $n$-limit,
the non-trivial exclusion for $m_a \lesssim 10^{-20}$~eV is obtained, somewhat stronger than constraints
obtained from the Lyman-$\alpha$ forest \citep{Haehnelt2017, Kobayashi2017}. Note that our bounds on ULAs
extend to much larger masses than the ones obtained from CMB and large-scale structure data in
\citep{Amendola2006, Marsh2010, Marsh2012, Hlozek2015, Hlozek2018}, but rely on the post-inflationary hypothesis.

The middle panel of \autoref{fig:ma-bound} shows the potential improvement corresponding to a sensitivity
$f_{\rm iso} < 0.067$ ($1\sigma$-level) for a future CMB s4 experiment. We observe that the exclusion limits
on $m_{a}$ become roughly one order of magnitude stronger. Finally, the right panel corresponds to the limit
$f_{\rm iso} < 0.016$ ($1\sigma$-level), potentially achievable with the combination of CMB s4 experiments
and advanced 21-cm observations based on an optimistic SKA2 configuration, which would lead to another order
of magnitude improvement in $m_{a}$. In the most optimistic case, a limit of $m_{a}\gtrsim 10^{-14}$~eV could
be achieved while a more robust bound (with respect to the mass-temperature dependence) is
$m_{a}\gtrsim \text{few}\times 10^{-19}$~eV.

\section{Discussion and conclusion}
\label{sec:conclusions}

In this paper, we investigated mass bounds on ALP dark matter derived from observations of the large-scale structure
of the Universe. If the PQ symmetry is broken after inflation, additional isocurvature fluctuations are generated on
top of the adiabatic spectrum. These isocurvature fluctuations have a white noise power spectrum, with a cutoff corresponding
to the size of the horizon at the time when the ALP field starts to oscillate. By requiring that the ALPs provide all DM,
we derived the amplitude of the isocurvature fluctuations relative to the adiabatic component as a function of the ALP
model parameters. In our analysis, we used the following model assumptions:
\begin{enumerate}[a)]
\item We assume a two-parameter model for the mass-temperature dependence of ALPs given by a power law with index $n$,
and a parameter $b$ that determines the temperature at which the zero-temperature mass is reached.
\item We use the harmonic approximation of the ALP potential and assume that the DM energy density is generated by the
realignment mechanism. This introduces uncertainties in our calculation of the DM density and the amplitude of isocurvature
fluctuations. However, the overall uncertainties are dominated by the  ALP model dependence in assumption $a)$.
\item The PQ symmetry is broken after the end of inflation. For very small ALP masses, this requires relatively high values
of the energy scale of inflation. Generically, this could give rise to tensor-to-scalar ratios in the observable range.
A more thorough discussion of this condition is presented in \autoref{sec:axions_topo}.
\item The cosmological evolution of ALPs can effectively be treated as cold DM and the imprint on large-scale structure
is determined by their initial power spectrum. For the ALP mass range and the cosmological scales considered here, this
assumption is well justified.
\end{enumerate}
With these assumptions, we were able to derive constraints on the ALP mass for different models of the mass-temperature
dependence. In particular, we focused on observations probing linear scales given by CMB and HI intensity mapping experiments.
Our findings are the following:
\begin{enumerate}[i)]
\item CMB observations by the Planck satellite set a lower bound on the ALP mass ranging from $10^{-20}$ eV for very shallow
mass dependences to $10^{-16}$ eV for very steep ones. The limits are generally tighter for small values of $b$.
\item CMB s4 experiments will improve these limits by roughly one order of magnitude due to the increased sensitivity in
polarization measurements at high multipoles.
\item Adding HI intensity mapping in the form of an SKA2-like experiment could further boost these limits by an additional
order of magnitude, providing a limit of $m_a\geq 10^{-14}$ eV in the most optimistic scenario.
\end{enumerate}

Our results are complementary to other works \citep{Haehnelt2017, Kobayashi2017,Amendola2006, Marsh2010, Marsh2012,
Hlozek2015, Hlozek2018}. Typically, our derived limits are stronger, but rely on the assumptions summarized above
(in particular, the post-inflationary scenario). While isocurvature constraints are already well established in the
pre-inflationary axion scenario, our analysis has shown that also in the post-inflationary case, isocurvature fluctuations
are a robust prediction of ALP DM models and lead to powerful constraints. Note that our results are based only on
gravitational effects and do not require assumptions about possible couplings to photons.

Concerning assumption b), let us mention that including the full ALP potential will give rise to an additional production
mechanism from cosmic strings. This may change the energy density and the amplitude of isocurvature fluctuations by factors
of order unity. However, it is expected that the main feature of the scenario, namely a white noise isocurvature power
spectrum at large distance scales, is a robust prediction.

Important questions in the future concern the cosmological evolution of miniclusters, especially in the nonlinear regime
of structure formation and for scales and masses where they cannot necessarily be treated as cold DM. While the observations
considered here focused on linear scales, nonlinear scales might provide an additional rich phenomenology for ALPs.

\acknowledgments This research was supported by the Excellence
Initiative of the German Federal and State Governments at Heidelberg
University, by the European Union's Horizon 2020 research and
innovation programme under the Marie Sklodowska-Curie grant agreement
No 674896 (Elusives), and by the Heidelberg Karlsruhe Research
Partnership (HEiKA).

\bibliographystyle{JHEP}
\bibliography{axion_cosmo}

\end{document}